\documentclass[preprint,12pt]{elsarticle}   

\usepackage{amssymb}
\usepackage{dsfont}
\usepackage{url}
\usepackage{amsmath}
\usepackage{amsthm}
\usepackage{booktabs}
\usepackage{siunitx}
\usepackage{adjustbox}
\usepackage{array}
\usepackage{url}
\usepackage{geometry}

\usepackage{xurl}                       % better URL breaks
\usepackage[hidelinks,breaklinks]{hyperref}

\usepackage{graphicx}
\usepackage{eurosym}
\usepackage{multicol}

\usepackage{url}
\usepackage{breakurl}
\usepackage{subcaption}
\usepackage{pdflscape}

\graphicspath{{doc/images/}}
\journal{Journal of Transport Geography}

\begin{document}

\begin{frontmatter}



\title{Metro 3 in Brussels under uncertainty: scenario-based public  transport accessibility analysis}

\author[label1,label3]{Brecht Verbeken\corref{cor1}}
\cortext[cor1]{Corresponding author.}
\ead{brecht.verbeken@vub.be}
\author[label1,label3]{Arne Vanhoyweghen}
\ead{arne.vanhoyweghen@vub.be}
\author[label1,label2,label3]{Vincent Ginis}
\ead{vincent.ginis@vub.be}
\affiliation[label1]{organization={Department Business Technology and Operations, Data Analytics Laboratory, Vrije Universiteit Brussel (VUB)},
           addressline={Pleinlaan 2},
           city={Brussels},
            postcode={1050},
            %state={},
           country={Belgium}}
\affiliation[label2]{organization={School of Engineering and Applied Sciences, Harvard University},
           city={Cambridge},
           postcode={02138},
           state={Massachusetts},
           country={USA}}

%\affiliation[label2]{organization={Department Business Technology and Operations, Data Analytics Laboratory, Vrije Universiteit Brussel (VUB)},
%            addressline={Pleinlaan 2},
%            city={Brussels},
%            postcode={1050},
%            country={Belgium}}

\affiliation[label3]{
  organization={imec-SMIT, Vrije Universiteit Brussel},
  addressline={Pleinlaan 9},
  city={Brussels},
  postcode={1050},
  country={Belgium}
}

%\author{Brando Vagenende, Brecht Verbeken, Marie-Anne Guerry} %% Author name

%\affiliation{organization={Department Business Technology and Operations, Data Analytics Laboratory, Vrije Universiteit Brussel (VUB)},
%            addressline={Pleinlaan 2},
%            city={Brussels},
%            postcode={1050},
%            country={Belgium}}

%% Abstract
\begin{abstract}
Metro Line 3 in Brussels is one of Europe's most debated infrastructure projects, marked by escalating costs, delays, and uncertainty over completion. Yet no public accessibility appraisal exists to inform this policy debate. This paper provides a scenario-based assessment of the spatial and distributional accessibility impacts of Metro 3 using schedule-based public transport data. Official GTFS feeds from all regional operators were combined and adapted to represent three scenarios: 
(i) the current network, (ii) partial implementation of Metro 3 (conversion of the southern premetro section), and (iii) full implementation to Bordet. Accessibility was measured as public transport travel time between 647 evenly spaced 500 m 
points covering the Brussels-Capital Region. Simulations were conducted for morning, evening,
and weekend midday periods, each with three departure times 
$(t-10,\quad t, \quad t+10)$,
capturing robustness to short-term timetable variation. Results show substantial but uneven accessibility gains, with the largest improvements occurring in neighborhoods with below-average incomes. 
Temporal robustness analysis highlights variability in accessibility outcomes, underscoring the need to account for uncertainty in departure timing. These findings suggest that Metro 3 has the potential to reduce socio-spatial inequalities in accessibility, providing transparent evidence for a project where public debate is dominated by cost concerns.
\end{abstract}


\begin{highlights}
\item GTFS-based simulation of Brussels public transport accessibility.
\item Three Metro 3 scenarios: status quo, half build, and full build.
\item Accessibility measured between 647 evenly spaced 500 m points.
\item Temporal robustness with multiple time slices and day types.
\item Accessibility improvements cluster in a subset of more population dense and disadvantaged neighborhoods.
\end{highlights}

%% Keywords
\samepage

\end{frontmatter}


\section{Introduction}

Metro Line~3 in Brussels, an envisioned north–south automated metro linking Forest (south) to Evere (north), has become one of the region’s most debated transport investments. Touted as a remedy for congestion and a boost to urban mobility, the project has been plagued by spiraling costs and delays. Initial cost estimates of \euro0.85 billion (2009) had risen to \euro2.3 billion by 2022; by 2023, projections stood at around \euro4.7 billion~\citep{Walker2024} , while the most recent scenario analyses estimate the total could reach nearly \euro5 billion or even rise to \euro7 billion under a public–private‑partnership model~\citep{Richards2025}.

Construction setbacks (notably at Palais du Midi and Brussels-North) and permit complications have repeatedly pushed timelines, amplifying concerns about feasibility and phasing~\citep{Richards2025,RTBF2025}. As of 2025, both scope and financing remain unsettled, with scenarios on the table ranging from a full build-out to a partial conversion or a halt~\citep{BRUZZ2025, Ferreira2025}.

In an internal report, the transit authority (STIB-MIVB) outlined options including a full build-out from Albert to Bordet (requiring an additional \euro3.6~billion investment) and a partial ``slimmed-down'' option converting only the existing pre-metro tram tunnel between Albert and Brussels-North (costing around \euro1~billion)~\citep{BRUZZ2025}. Despite the magnitude of Metro~3’s impacts and costs, no public cost--benefit analysis or accessibility appraisal has been released to justify the project’s socio-economic merit. Indeed, some policymakers have argued that an updated assessment of benefits is urgently needed given that the line’s cost ``is much more expensive than initially estimated'' and circumstances such as teleworking patterns have changed~\citep{DeStandaard2024}. In conclusion no published, reproducible analysis quantifies Metro 3’s accessibility impacts.

Within transport geography, this study is situated in the literature emphasizing accessibility as a key metric for evaluating transport and land-use projects. Accessibility, the ease of reaching valued opportunities, provides a people-centered perspective on infrastructure impacts, in contrast to narrow mobility indicators ~\citep{Geurs2004}. 

Following this paradigm, many scholars have called for major projects like new transit lines to be judged by their contributions to accessibility and spatial equity rather than just travel time savings or ridership. In the context of Brussels and Belgium, several studies have mapped and analyzed baseline accessibility patterns. For example, Vandenbulcke et al. ~\citep{Vandenbulcke2009} compared accessibility structures across Belgium and found significant regional disparities. Focusing on the capital, Lebrun~\citep{Lebrun2018} measured public transport travel times to key activity centers in Brussels and revealed pronounced intra-urban heterogeneity: roughly 30\% of the Brussels population live in neighborhoods with poor transit accessibility to jobs and services. In Brussels, where lower-income districts have lower car ownership and higher public-transport dependence, project appraisal should consider whether new lines deliver clear travel-time reductions in these areas,beyond improvements in citywide averages. However, to date no study has specifically examined how Metro~3 would reshape the accessibility landscape of Brussels. The present work aims to fill this gap by providing an accessibility-focused evaluation of Metro~3’s potential impacts, thereby contributing to the broader discourse on urban spatial inequality and transport justice~\citep{Martens2017, Pereira2017}.

Methodologically, our study builds on recent advances in accessibility simulation using General Transit Feed Specification (GTFS) data. The proliferation of GTFS (standardized public transport schedules) and open-source routing engines now enables detailed, schedule-based accessibility analyses for both existing and planned transit networks. Researchers have demonstrated the value of GTFS-driven analysis for transit planning: for instance, Conway et al.~\citep{conway2018} developed a public transport sketch planning tool that accounts for uncertainty in travel times, while Pereira et al.~\citep{Pereira2019} employed GTFS-based cumulative opportunity measures to evaluate future Bus Rapid Transit scenarios in Rio de Janeiro. Such tools make it possible to construct ``counterfactual'' schedule scenarios and compare accessibility outcomes across different network configurations. 

In this study, we leverage GTFS-based routing to simulate three scenarios for Metro~3 within the broader framework of existing public transport in the Brussels-Capital Region:
\begin{enumerate}
    \item \emph{status quo} scenario with no metro (the existing tram corridor retained)
    \item \emph{partial conversion} scenario (trams replaced by metro between Albert and Brussels-North, but no extension beyond Brussels-North)
    \item \emph{full metro} scenario (metro line running from Albert through the new tunnel to Bordet)
\end{enumerate}

 By generating GTFS feeds for each scenario and computing multi-modal travel times, we can quantify how Metro~3’s partial or full implementation would change access to opportunities across the Brussels metropolitan area. To our knowledge, this is one of the first applications of GTFS-driven accessibility modeling for an unreleased major transit project in Europe, and it demonstrates a novel approach to ex-ante transit appraisal.

A further innovation of our approach is the consideration of \textit{temporal robustness} in accessibility outcomes, informed by the concept of ergodicity breaking. Typically, accessibility studies evaluate a single departure time (often the morning peak hour) as representative of network performance. In reality, however, travel times and transit service levels vary across the day and week, meaning a one-shot temporal sample may not reflect a traveller’s everyday experience. Crucially, accessibility is generated by a non-ergodic process: the time-average along an individual’s trajectory generally diverges from the ensemble snapshot at an arbitrary instant. A single “representative” departure therefore estimates the wrong object for welfare and behavior; our multi-time, multi-day sampling instead approximates the relevant time-average and exposes downside risk masked by snapshots~\citep{Peters2019, Vanhoyweghen2025Ergodicity}.

In non-ergodic systems, the time-average experience of an individual can differ markedly from the ensemble average at an arbitrary moment. Concretely, we calculate accessibility at three departure times in the morning peak (the scheduled time $t$ and $t \pm 10$~min) and across three representative day-types (weekday, evening, and Saturday) to capture variability in service schedules. By evaluating nine time/day combinations per scenario, we obtain a distribution of accessibility outcomes that reflects typical fluctuations.

By quantifying accessibility under three realistic scenarios, this study offers the first reproducible evidence of Metro 3’s impacts, translating technical planning assumptions into implications that matter for both decision-makers and citizens.

In summary, this paper addresses three key gaps in the literature. First, it presents the first accessibility analysis of the proposed Brussels Metro~3, offering evidence on the project’s spatial impacts in the absence of any official public appraisal. Second, it advances the use of GTFS-based scenario modeling for transport infrastructure evaluation, illustrating how future ``what-if'' transit networks can be rigorously assessed using open data and reproducible methods. Such counterfactual accessibility modeling remains relatively uncommon in transport geography, and our case study demonstrates its potential value for ex-ante policy analysis~\citep{Pereira2019, painter2018innovation}. Third, it introduces a temporal robustness perspective into accessibility analysis, addressing the limited attention of departure-time variability in most accessibility studies~\citep{conway2018, Stepniak2019,boisjoly2016daily}. By doing so, we respond to recent calls for more resilient and temporally aware accessibility metrics that better reflect travelers’ lived experiences. Collectively, these contributions aim to enrich both the empirical understanding of Metro~3’s implications and the methodological toolbox for transport geographers and planners.

The remainder of this article is structured as follows. Section~2 describes the study context and data, including the construction of GTFS scenarios for the Metro~3 options. Section~3 details the methodology, including accessibility measures, the multi-temporal simulation procedure, and analytical metrics for comparing scenarios. Section~4 presents the results of the accessibility simulations, highlighting the spatial distribution of benefits under each Metro~3 scenario and the sensitivity to temporal variability. Finally, Section~5 discusses the implications of these findings as well as the limitations of our approach, we conclude with a summary of key insights and recommendations for policy and future research.

\section{Study area, data, and scenarios}

The study area is the \emph{Brussels–Capital Region (BCR)}. We compute public-transport travel times on the full multimodal network (\emph{metro, tram, bus, and rail}) by integrating the official GTFS feeds for \emph{STIB/MIVB}, \emph{De Lijn}, \emph{TEC}, and \emph{SNCB/NMBS}~\citep{stib_gtfs,delijn_gtfs,tec_gtfs,sncb_gtfs}. Rail is retained both because the suburban Brussels S Train provides dense intra-urban stopping patterns within the BCR and because commuting flows from the wider metropolitan area interact with intra-BCR accessibility.

Simulations are anchored to three \emph{representative 2025 service days} and, for each, three departure instants to capture short-horizon timetable sensitivity:
\begin{itemize}\setlength\itemsep{0.2em}
    \item \textbf{Weekday AM peak:} Tue 10 June 2025 at 07:50, 08:00, 08:10.
    \item \textbf{Weekday PM peak:} Thu 12 June 2025 at 17:20, 17:30, 17:40.
    \item \textbf{Saturday mid-day:} Sat 12 July 2025 at 12:50, 13:00, 13:10.
\end{itemize}
These slices implement the canonical peak/off-peak bands used in schedule-based accessibility studies; the \emph{partial} and \emph{full} Metro~3 scenarios necessarily adapt the STIB/MIVB feed relative to these base calendars (details in \S\ref{sec:scenario_construction}). Evaluating $t{-}10$, $t$, and $t{+}10$ per band operationalises the paper’s temporal robustness principle (cf.~\S1).

\subsection{Spatial sampling, OD construction and data sources}
Origins and destinations are the centres of 647 tiles on a regular $500{\times}500$\,m lattice clipped to the BCR. This resolution balances coverage and tractability. The expected straight-line distance from a random point in a $500$\,m square to its centre is $\approx 191$\,m; at typical urban walking speeds ($\approx 1.3$\,m/s), this corresponds to $\approx 2.5$\,min. Hence the origin-distance (OD) paths between tile centres approximate real journeys with an \emph{expected} access$+$egress overhead of $\approx 5$\,min, delivering citywide coverage while keeping the OD matrix computationally feasible.

We use four GTFS feeds (all accessed 2024--06--10): STIB/MIVB, De Lijn, TEC, and SNCB/NMBS~\citep{stib_gtfs,delijn_gtfs,tec_gtfs,sncb_gtfs}. The pedestrian network for access/egress is derived from OpenStreetMap (Geofabrik extract \\ \texttt{belgium-latest.osm.pbf}) \citep{osm_contributors_2025, geofabrik_belgium_2025}. Feeds are loaded into a single multimodal graph with \texttt{r5py}/R5~\citep{fink2022r5py}, enabling inter-agency transfers at co-located stops. Following established schedule-based accessibility practice~\citep{conway2017evidence,conway2018,conway2019getting}, we evaluate \emph{earliest arrival time}.

\subsection{Scenario construction (status quo, partial, full)}
\label{sec:scenario_construction}
This study evaluates three network configurations: (i) \emph{status quo} (premetro retained), (ii) \emph{partial conversion} (Albert--Brussels-North converted to metro), and (iii) \emph{full Metro~3} (Albert--Bordet). Each scenario is implemented as a derived GTFS feed obtained by editing the base feeds for the 2025 slices above.

Our primary source for scenario design was the official Metro~3 project site~\citep{metro3_official}. Public communication is sparse and occasionally inconsistent, so we complemented the published material with limited engineering inference to produce internally coherent GTFS scenario feeds. \\
\emph{Full Metro 3.} We encoded the line using the stations published by the project, placed at the coordinates communicated in official materials: Albert, Horta, Parvis de Saint-Gilles, Porte de Hal, Gare du Midi, Toots Thielemans, Anneessens, Bourse, De Brouckère, Rogier, Gare du Nord (Brussels-North), Liedts, Colignon, Verboekhoven, Riga, Tilleul, Paix, Bordet. Inter-station run times were derived from great-circle distances and calibrated against performance on existing Brussels metro lines, yielding an end-to-end runtime just under 20 minutes, in line with official statements~\citep{metro3_official}. Because the tram network is slated to be restructured if Metro~3 opens, we also reflected the operator’s patterns in this scenario: Tram~7 is extended from Vanderkindere to Albert; Tram~4 is curtailed to Stalle--Albert; and Tram~10 (successor to the former Tram~3) is adjusted to operate between Esplanade and Rogier along its published alignment. For trams, inter-station run times were taken from observed segment times on the corresponding existing lines. Service frequency (headways and daily runs) was replicated from current STIB/MIVB timetables for those lines. \\

\emph{Partial conversion (Albert--Brussels-North).} We implemented metro service only between Albert and Brussels-North (Albert, Horta, Parvis de Saint-Gilles, Porte de Hal, Gare du Midi,Toots Thielemans, Anneessens, Bourse, De Brouckère, Rogier, Gare du Nord (Brussels-North)), using the same dwell and run-time assumptions as on the corresponding segment of the full line.In line with the operator’s public remarks about interim operational constraints, including the absence of a local depot on this line, we reduced metro supply by approximately 20\% on the affected corridor in this scenario~\citep{BRUZZ2025}.A key constraint is the absence of a dedicated depot. Operationally, rolling stock must be positioned from elsewhere each day, which limits staging and turnbacks and caps the maximum number of trainsets that can be in service simultaneously. Trams~4, 7, and 10 follow the same routing patterns as in the full-line scenario. \\

\emph{Computation} OD travel times between all tile pairs are computed for every scenario$\times$time slice using batched \texttt{r5py}/R5 runs . Computations ran on the Vlaams Supercomputer Centrum (VSC); cumulative workload amounted to $\approx$\,\textit{4 CPU-years}.

\section{Methods: data analysis, robustness, and scenario comparison}
\label{sec:methods}

\subsection{Raw simulation outputs and panel structure}
For each scenario $s\in\{\textit{baseline},\textit{partial conversion},\textit{full metro 3}\}$, for each day-type $d\in\{\text{AM-peak},\text{PM-peak},\text{Sat}\}$, and for each departure instant $k\in\{-10,0,+10\}$ within that day-type, the routing engine returns directed door-to-door public-transport travel times between spatial samples (tile centroids) $i,j\in\mathcal{S}$:
\[
T_{s,d,k}(i,j)\in\mathbb{R}_{\ge 0}\cup\{\mathrm{NA}\},\qquad i,j=1,\dots,N.
\]
Rows in the analysis table have the schema
\[
(o_i,d_j,s,\texttt{snapshot\_time}=k,\; \texttt{time\_s}=T_{s,d,k}(i,j),\; \texttt{day},\; \texttt{start\_time}),
\]
e.g.\ 
\texttt{origin}=(4.2493,50.8195), \texttt{destination}=(4.2564,50.8195), \texttt{scenario}=\newline \texttt{baseline}, \texttt{snapshot\_time}=\texttt{t-10}, \texttt{time\_s}=3119, \texttt{day}=\texttt{0610}, \texttt{start\_time}=\texttt{0750}.
Eight spatial samples were structurally unreachable (six in Laeken Park, one within NATO grounds in Evere, and one between the tracks at Bruxelles-Midi), so we drop them as both origins and destinations. This yields $N=639$ nodes and $M=N(N-1)=639\times 638$ directed OD pairs per slice, for $K=27$ slices overall, i.e.\ $M\times K$ rows in the long panel.

\subsection{Distributional diagnostics: quantiles, shift functions, and 
  \texorpdfstring{$\Delta$ECDF}{ΔECDF}}
Rather than only reporting a single mean effect, we characterise how \emph{the entire distribution} of OD times shifts under each scenario, using three complementary, nonparametric diagnostics:

\paragraph{(i) Deciles of $\Delta T$}
For each comparison $s$, we report the deciles of $\Delta T_s$ across OD pairs $(i,j)\in\mathcal S$:
\begin{equation*}
Q^{(\Delta)}_{s}(p)
= \operatorname{Quantile}\!\left(\left\{\Delta T_s(i,j)\,:\,(i,j)\in\mathcal{S}\right\},\, p\right),
\qquad p\in\{0,0.1,\dots,1\}.
\end{equation*}
This shows heterogeneity: e.g.\ whether the $90^{\mathrm{th}}$ percentile OD pair still improves. We also compute key shares (in $\%$):
\begin{align*}
\text{share}_{\text{improved}}
&= \frac{\#\{(i,j)\in\mathcal{S} : \Delta T_s(i,j) < 0\}}{|\mathcal{S}|}, \\[0.75em]
\text{share}_{\geq 5\%\text{ better}}
&= \frac{\#\left\{(i,j)\in\mathcal{S} : \tfrac{\Delta T_s(i,j)}{\tilde T_{\text{baseline}}(i,j)} \leq -0.05\right\}}{|\mathcal{S}|}.
\end{align*}
Here, $\#\{\cdot\}$ denotes the cardinality (number of elements) of the set.

\begin{table}[htbp]

\centering
\scriptsize
\begin{adjustbox}{max width=\textwidth}
\begin{tabular}{
  l
  S
  S
  S
  S
  S
  S
}
\toprule
comparison 
& {mean $\Delta$t (s)} 
& {median $\Delta$t (s)} 
& \multicolumn{1}{c}{improved (\%)} 
& \multicolumn{1}{c}{equal$\pm$1s (\%)} 
& \multicolumn{1}{c}{$\geq$5\% better (\%)} 
& \multicolumn{1}{c}{$\geq$10\% better (\%)} \\
\midrule
partial conversion 
& -16.5 & 0
& 49.8 & 28.8
& 5.6 & 0.7 \\
\addlinespace
full metro 3 
& -36.1 & -6.1
& 52.8 & 28.1
& 9.4 & 2.8 \\
\bottomrule
\end{tabular}
\end{adjustbox}
\caption{Summary statistics of travel time changes relative to the status quo. 
Partial conversion yields modest mean savings and limited shares of substantial improvements, 
while the full metro~3 delivers larger average reductions and a greater proportion of OD pairs with improvements exceeding 5\% and 10\%.}

\label{tab:summary}
\end{table}

\paragraph{(ii) Shift function of aggregated times} Let $A=\{\tilde T_{\text{baseline}}(i,j):(i,j)\in\mathcal{S}\}$ and $B=\{\tilde T_{s}(i,j):(i,j)\in\mathcal{S}\}$. The \emph{shift function} compares quantiles of $B$ and $A$ across $p\in(0,1)$:
\[
\Delta Q_s(p)\;=\;Q_B(p)\;-\;Q_A(p),\qquad Q_X(p)=\operatorname{quantile}(X,\,p).
\]
Unlike mean differences, $\Delta Q_s(p)$ exposes where in the distribution the scenario helps or harms (e.g.\ large gains for slow OD pairs but little change for already fast pairs). This is standard in robust distributional comparisons~\citep{Wilcox2012}.

\paragraph{(iii) Difference of empirical CDFs ($\Delta$ECDF)}
With empirical CDFs
\begin{align*}
\widehat F_A(t)
&= \frac{\#\{(i,j)\in\mathcal{S} : \tilde T_{\text{baseline}}(i,j)\le t\}}{|\mathcal{S}|}, \\[0.75em]
\widehat F_B(t)
&= \frac{\#\{(i,j)\in\mathcal{S} : \tilde T_{s}(i,j)\le t\}}{|\mathcal{S}|}.
\end{align*}

we plot
\[
\Delta \widehat F_s(t)\;=\;\widehat F_B(t)-\widehat F_A(t).
\]
Positive values at a given $t$ mean a larger share of OD pairs can be completed in $\le t$ seconds under the scenario. This is closely related to the Kolmogorov–Smirnov perspective on distributional change~\citep{Handcock1999,Massey1951}.

\begin{table}[htbp]
\centering
\scriptsize
\begin{adjustbox}{max width=\textwidth}
\begin{tabular}{
  l
  S
  S
  S
  S
  S
  S
}
\toprule
comparison 
& {p0 $\Delta$t (s)} & {p20 $\Delta$t (s)} & {p40 $\Delta$t (s)} 
& {p60 $\Delta$t (s)} & {p80 $\Delta$t (s)} & {p100 $\Delta$t (s)} \\
\midrule
partial conversion 
& -693 & -76 & -20
& 0 & 40 & 551 \\
\addlinespace
full metro 3 
& -1703 & -93 & -29 
& 0 & 35 & 567 \\
\bottomrule
\end{tabular}
\end{adjustbox}
\caption{Percentile distribution of travel time changes ($\Delta$t, in seconds) relative to the status quo. 
The full metro~3 scenario exhibits more extreme improvements in the lower tail while both scenarios show similar upper-tail increases; see also Figure~\ref{fig:decdf}.}
\label{tab:percentiles}
\end{table}

\subsection{Spatial visualisation of per-origin time benefits}
\label{sec:maps}

\paragraph{From OD panels to per-origin outbound changes}
For each scenario \(s\in\{\text{partial},\text{full}\}\) we summarise the \emph{outbound} accessibility of an origin \(i\) by the average of OD times to all destinations:
\begin{align*}
\bar T_{i,s}
&= \frac{1}{M_i}\sum_{\substack{j=1 \\ j\neq i}}^{N}\overline{T}_{s}(i,j),
\quad
M_i = \#\{\,j\neq i : \overline{T}_{s}(i,j)\in\mathbb{R}\,\}.
\end{align*}

Here \(\overline{T}_{s}(i,j)\) is the \emph{mean} over slices
\[
\overline{T}_{s}(i,j)\;=\;\frac{1}{K}\sum_{k\in\mathcal{K}} T_{s,k}(i,j),
\]
with \(K=3\) and \(\mathcal{K}=\{-10,0,+10\}\).
We then define the \emph{per-origin time change} (seconds) relative to the status quo:
\[
\Delta^{\mathrm{out}}_{i,s}\;=\;\bar T_{i,s}\;-\;\bar T_{i,\text{baseline}}\,,\qquad
\Delta^{\mathrm{out}}_{i,s}<0\ \Rightarrow\ \text{faster (benefit)}.
\]

An analogous day-specific version aggregates only within day-type/date \(d\):
\begin{align}
\overline{T}_{s,d}(i,j) &= \frac{1}{K}\sum_{k\in\mathcal{K}} T_{s,d,k}(i,j), \\[4pt]
\bar T^{(d)}_{i,s} &= \frac{1}{M^{(d)}_i} \sum_{j\neq i} \overline{T}_{s,d}(i,j), \\[4pt]
\Delta^{\mathrm{out}}_{i,s}(d) &= \bar T^{(d)}_{i,s} - \bar T^{(d)}_{i,\text{baseline}}.
\end{align}
Practically, the code groups the long table by  \texttt{(scenario, origin,} \newline \texttt{destination)}, collapses to the mean over slices, then averages across destinations to get $\bar T_{i,s}$; deltas $\Delta^{\mathrm{out}}_{i,s}$ are computed on the \emph{common origin support} with the baseline (inner merge).

\paragraph{Per-origin outbound variability (\(\Delta\) SD) maps}
In addition to mean-time changes, we map changes in \emph{temporal dispersion}. For each \((i,s)\) we take the population variance across slices at the OD level and aggregate as

\begin{align}
\operatorname{Var}_{s,d}(i,j) 
  &= \operatorname{Var}_{k\in\mathcal{K}}\!\bigl[T_{s,d,k}(i,j)\bigr] 
  \quad (\text{OD pair variability, day } d), \\[6pt]
V^{(d)}_{i,s} 
  &= \frac{1}{M^{(d)}_i}\sum_{j\neq i}\operatorname{Var}_{s,d}(i,j) 
  \quad (\text{avg. OD variability, day } d), \\[6pt]
\bar V_{i,s} 
  &= \frac{1}{|D|}\sum_{d\in D} V^{(d)}_{i,s} 
  \quad (\text{avg. OD variability across days}), \\[6pt]
\sigma_{i,s} 
  &= \sqrt{\bar V_{i,s}} 
  \quad (\text{st.dev. of avg. variability}).
\end{align}

\begin{align}
\Delta^{\mathrm{SD}}_{i,s}
  &= \sigma_{i,s}-\sigma_{i,\text{baseline}}
  \quad\text{(seconds; }\Delta^{\mathrm{SD}}_{i,s}<0 \Rightarrow \text{less variability).}
\end{align}

Values are computed on the \emph{common origin support}. We report the variance only at the scenario level, since each scenario provides nine data points, whereas day-wise comparisons rely on only three data points and thus offer limited statistical interpretability.

\paragraph{Grid geometry}
Origins are the centroids of a $500{\times}500$\,m regular lattice clipped to the BCR, yielding square tiles via
\[
\mathcal{P}_i=\big\{(x,y): |x-x_i|\le 250\ \wedge\ |y-y_i|\le 250\big\}.
\]
Expected centres without a nearby observed origin (within $35$\,m) are flagged as grid \emph{holes}. These holes are rendered as black squares on maps to make data gaps legible (the eight unreachable samples: Laeken Park~$\times 6$, NATO Evere, track interstice at Bruxelles-Midi). We associate the corresponding $\Delta^{\mathrm{out}}_{i,s}$ with each tile and color the tiles according to this value. Commune boundaries are overlaid for orientation only.

\paragraph{Map outputs.}
We render (i) an \emph{all-dates} map per scenario using $\Delta^{\mathrm{out}}_{i,s}$, and (ii) per-daycode maps using $\Delta^{\mathrm{out}}_{i,s}(d)$ for each snapshot date (labels derived from \texttt{MMDD} codes).

\subsection{Directionality diagnostics and justification for outbound focus}
\label{sec:directionality}

\paragraph{Matrix setup and decomposition}
For each scenario $s$, let $A_s\in\mathbb{R}^{N\times N}$ be the directed OD travel-time matrix returned by the simulator, with $(A_s)_{ij}$ the door-to-door public transport time (seconds) from origin $i$ to destination $j$ on the $N=639$ grid nodes (eight unreachable samples dropped). We decompose
\begin{align}
A_s &= S_s + K_s, \\
S_s &= \tfrac{1}{2}\bigl(A_s + A_s^\top\bigr), \quad &&\text{(symmetric)}, \\
K_s &= \tfrac{1}{2}\bigl(A_s - A_s^\top\bigr), \quad &&\text{(skew-symmetric)}.
\end{align}

which are orthogonal under the Frobenius inner product (so $\|A_s\|_F^2=\|S_s\|_F^2+\|K_s\|_F^2$). $S_s$ captures direction-free structure; $K_s$ captures pure directionality ($i{\to}j$ vs $j{\to}i$).

\paragraph{Directionality Index and pairwise thresholds.}
We quantify system-wide asymmetry by the scale-free
\[
\mathrm{DI}(s)\;=\;\frac{\|K_s\|_F}{\|S_s\|_F}\,,
\]
We obtained a DI of around $0.035$, which implies that 
directionality only explains
\[
\frac{\|K_s\|_F^2}{\|A_s\|_F^2}=\frac{\mathrm{DI}(s)^2}{1+\mathrm{DI}(s)^2}\approx 0.12\%
\]
of OD variance. I.e., the OD matrix is \emph{nearly symmetric}. Accordingly, we report outbound results only: with $\mathrm{DI}(s)=0.035$ (so directionality explains $\approx0.12\%$ of $|A_s|_F^2$), inbound patterns are effectively identical up to the small skew component $K_s$.

\begin{landscape}
\thispagestyle{empty}

\begin{figure}[p]
  \centering

  % Row 1
  \begin{subfigure}[t]{0.48\linewidth}
    \includegraphics[width=\linewidth]{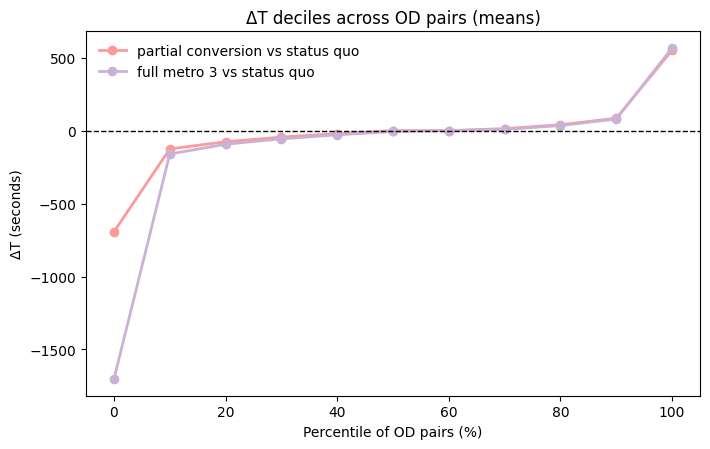}
    \caption{\textbf{Deciles of OD-pair travel time differences.} For each OD pair we compute $\Delta T = T_{\text{scenario}} - T_{\text{baseline}}$, then summarize the distribution of these differences across all OD pairs by its deciles. Negative values denote time savings. This captures the distribution of \emph{pairwise changes} in travel times.}
    \label{fig:quantiles}
  \end{subfigure}\hfill
  \begin{subfigure}[t]{0.48\linewidth}
    \includegraphics[width=\linewidth]{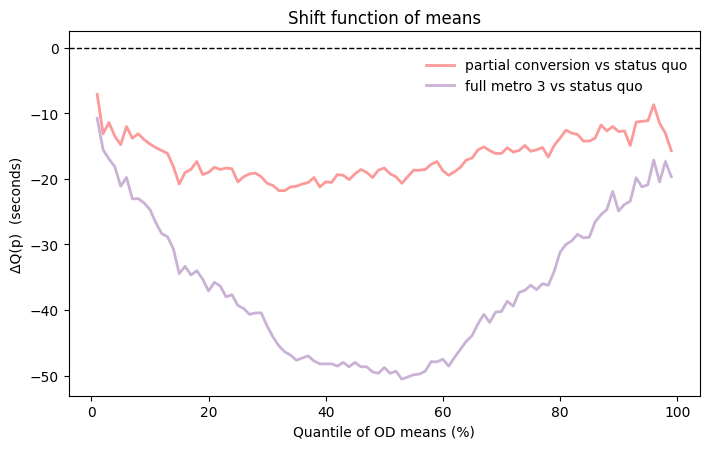}
    \caption{\textbf{Shift function of OD time distributions.} For the entire distribution of OD times under each scenario, we compute the quantile function $Q(p)$ of aggregated travel times and then plot $\Delta Q(p) = Q_{\text{scenario}}(p) - Q_{\text{baseline}}(p)$ over $p\in[1,99]$. Negative values denote time savings.}
    \label{fig:shift}
  \end{subfigure}

  \vspace{0.75em}

  % Row 2
  \begin{subfigure}[t]{0.48\linewidth}
    \includegraphics[width=\linewidth]{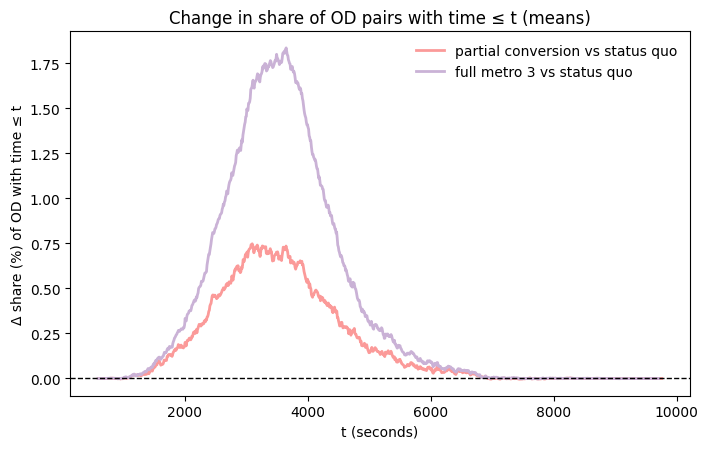}
    \caption{\textbf{Change in share of OD pairs completed within $t$.} $\Delta \widehat F(t)=\widehat F_{\text{scenario}}(t)-\widehat F_{\text{baseline}}(t)$, plotted in percentage points. Positive values mean more OD pairs are reachable within $t$.}
    \label{fig:decdf}
  \end{subfigure}\hfill
  \begin{subfigure}[t]{0.48\linewidth}
    \includegraphics[width=\linewidth]{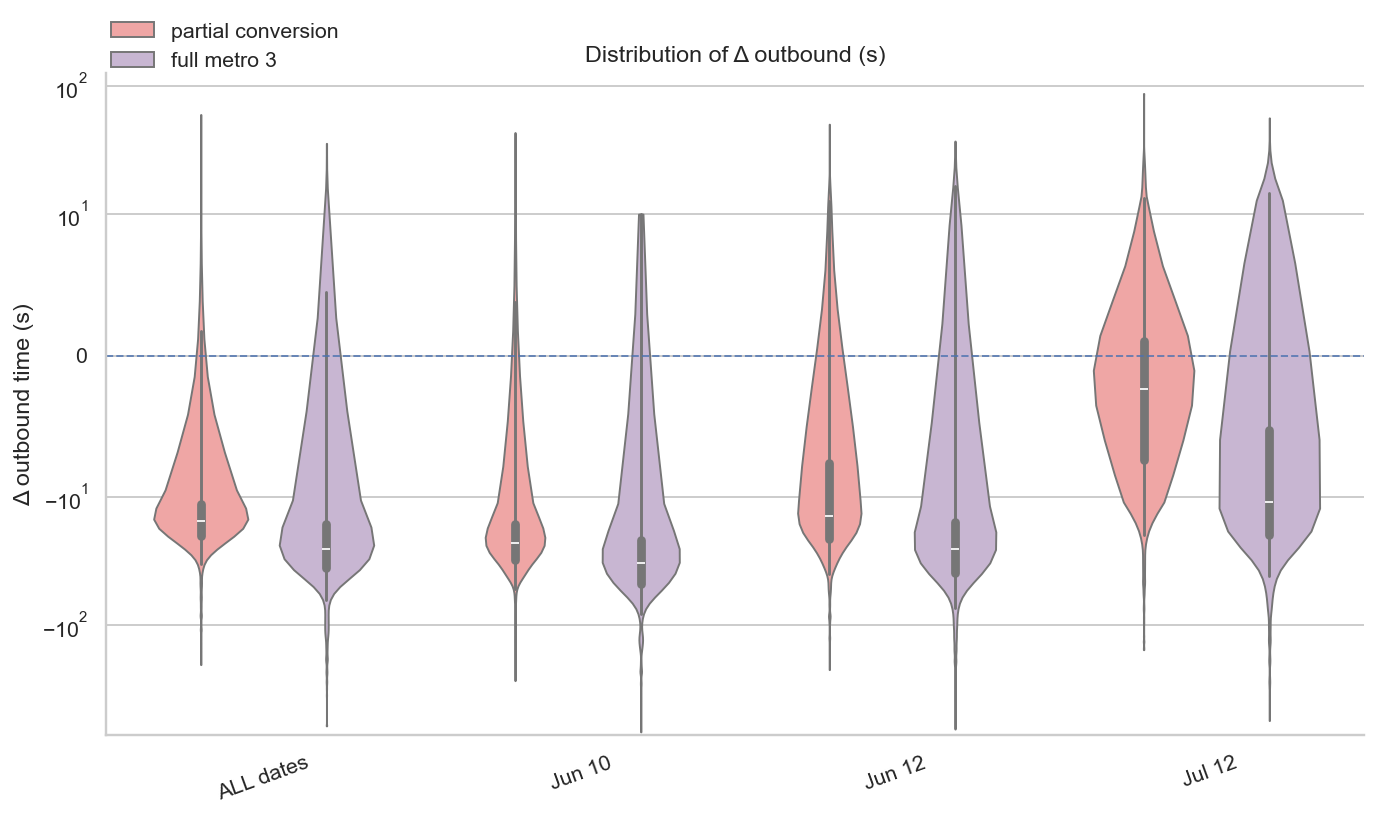}
    \caption{\textbf{Distribution of per-origin $\Delta$ outbound time.} Violin plots of $\Delta^{\mathrm{out}}=\bar T_{\text{scenario}}-\bar T_{\text{baseline}}$. Symmetric log $y$-scale (linear within $\pm 10$\,s).}
    \label{fig:distros}
  \end{subfigure}

  \caption{\textbf{Distributional diagnostics for time benefits.} Panels (a)–(b) use common-support OD pairs and scenario-aggregated OD times; negative values indicate faster travel. Together, they show broad, distribution-wide improvements for the full line and smaller gains for the partial conversion; panel (c) shows the difference in reachability within time $t$;  panel (d) summarises the same effects at the per-origin level.}
  \label{fig:overview-panels}
\end{figure}

\end{landscape}

% Page 1: priority maps
% ======================
\clearpage
\newgeometry{left=1cm,right=1cm,top=1cm,bottom=1cm} % whatever margins you want
\begin{landscape}
\thispagestyle{empty}
\begin{figure}[p]
  \centering
  \begin{subfigure}[t]{0.49\linewidth}
    \centering
    \includegraphics[width=\linewidth]{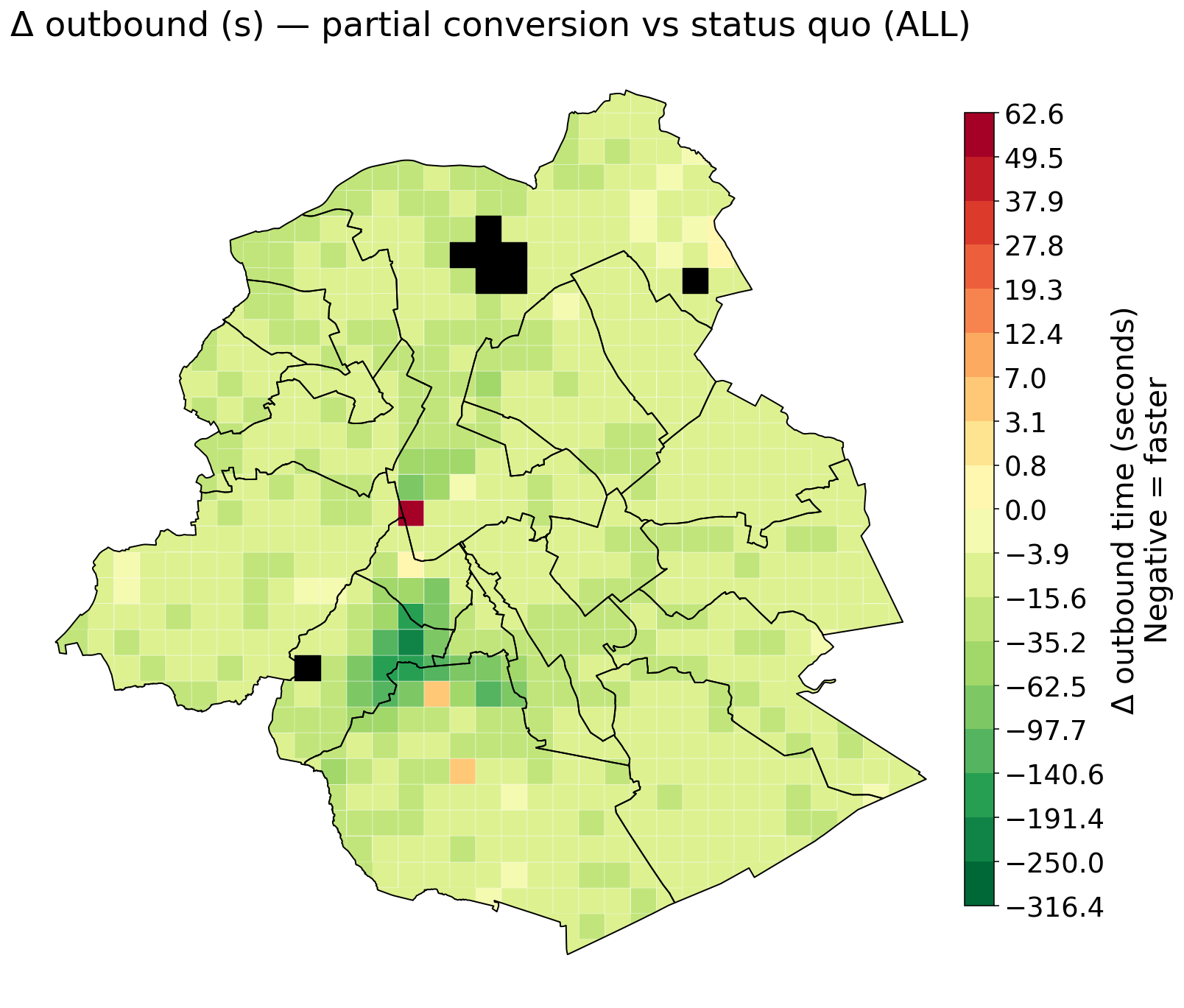}
    \caption{\textbf{All-days composite — partial conversion}}
    \label{fig:mapall-partial}
  \end{subfigure}\hfill
  \begin{subfigure}[t]{0.49\linewidth}
    \centering
    \includegraphics[width=\linewidth]{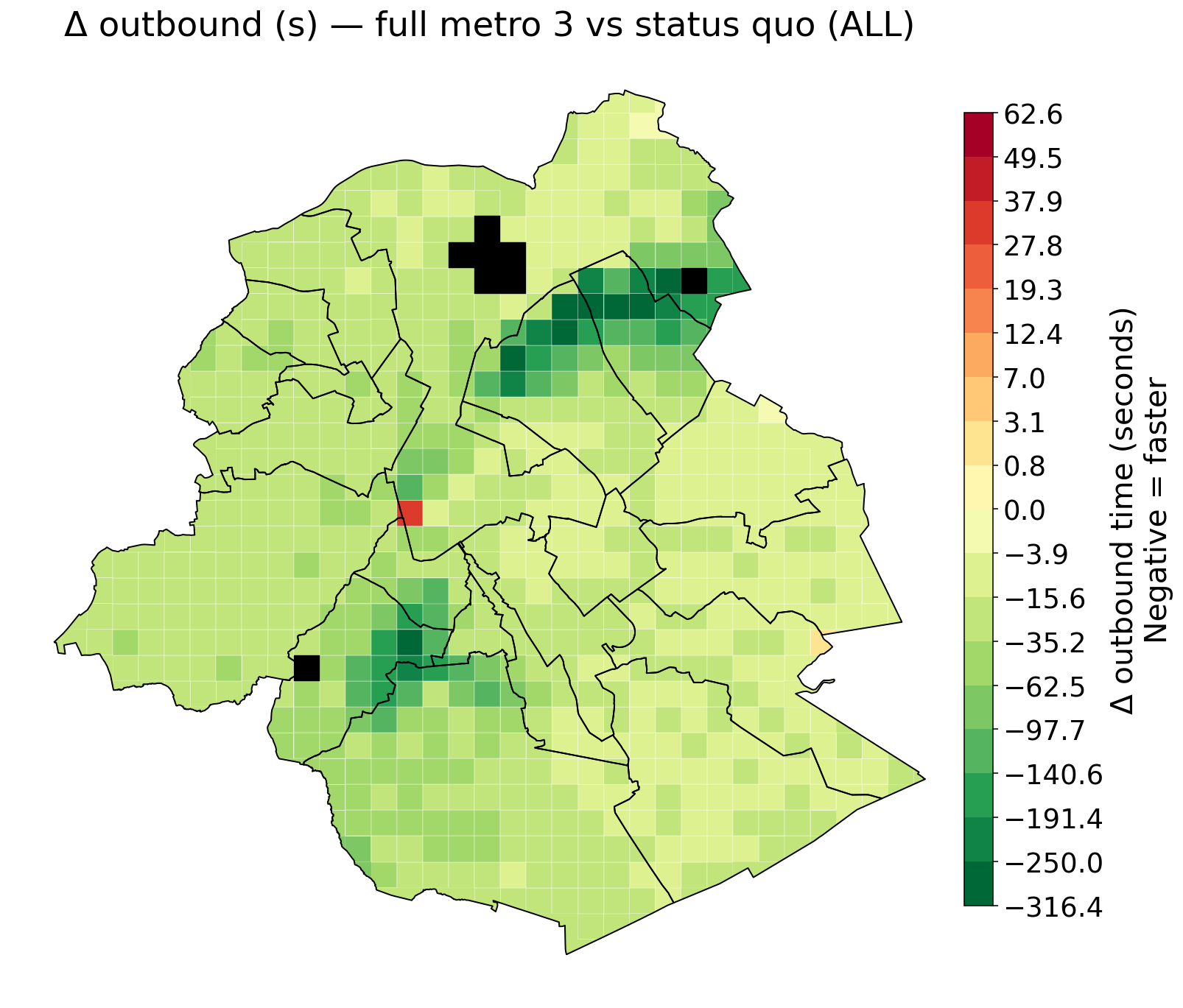}
    \caption{\textbf{All-days composite — full line}}
    \label{fig:mapall-full}
  \end{subfigure}

  \caption{\textbf{Systemwide maps.} All-days composites of differences in means for partial conversion and full line.}
  \label{fig:maps-priority}
\end{figure}
\end{landscape}
\restoregeometry

\clearpage

% =========================
% Page 2: dated 3×2 mosaic
% =========================
\clearpage
\newgeometry{left=1cm,right=1cm,top=1cm,bottom=1cm} % tighter margins for this page
\begin{landscape}
\thispagestyle{empty}
\begin{figure}[p]
  \centering

  % Row 1
  \begin{subfigure}[t]{0.32\linewidth}
    \centering
    \includegraphics[width=\linewidth]{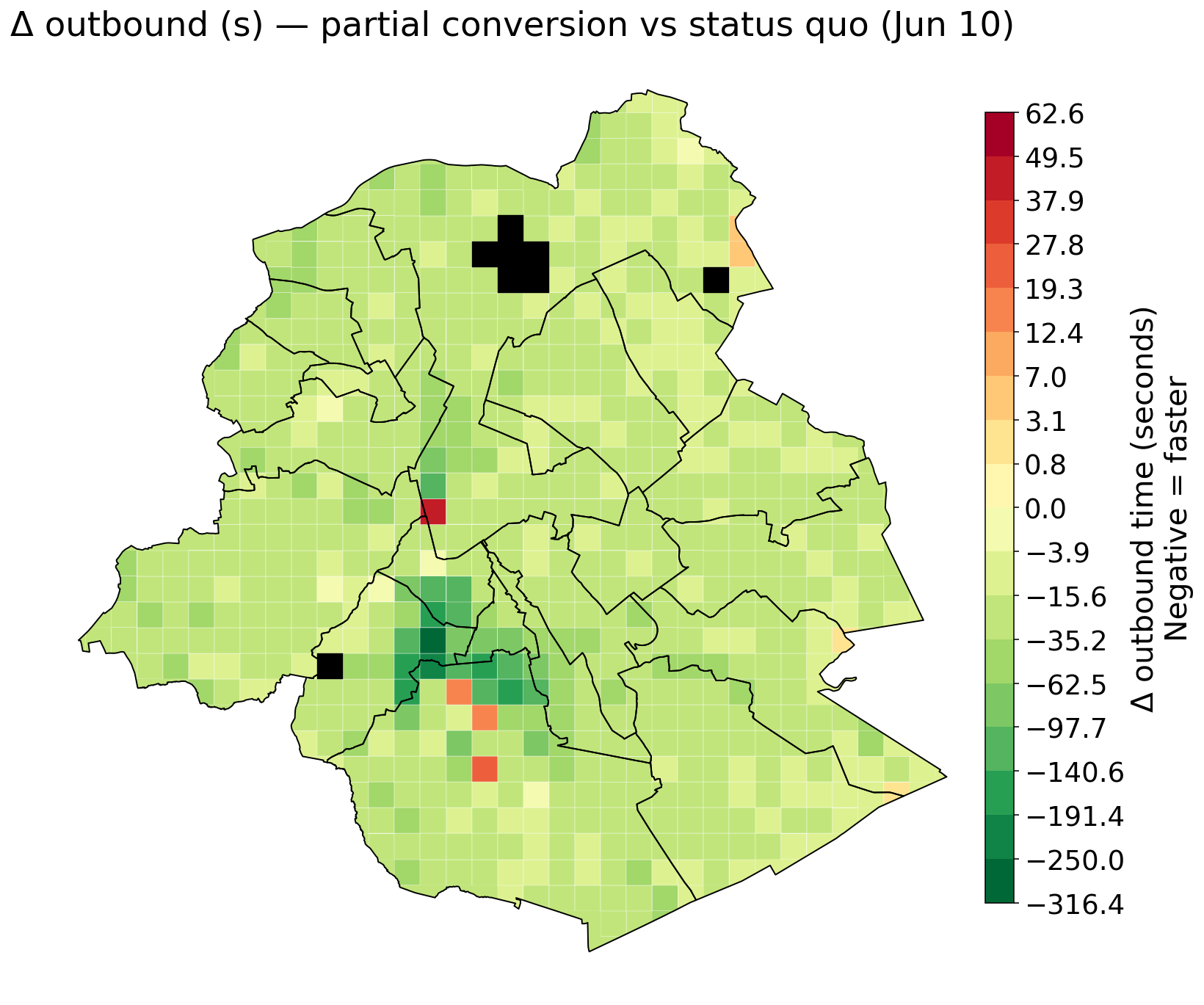}
    \caption{0610 — partial}
    \label{fig:map0610-partial}
  \end{subfigure}\hfill
  \begin{subfigure}[t]{0.32\linewidth}
    \centering
    \includegraphics[width=\linewidth]{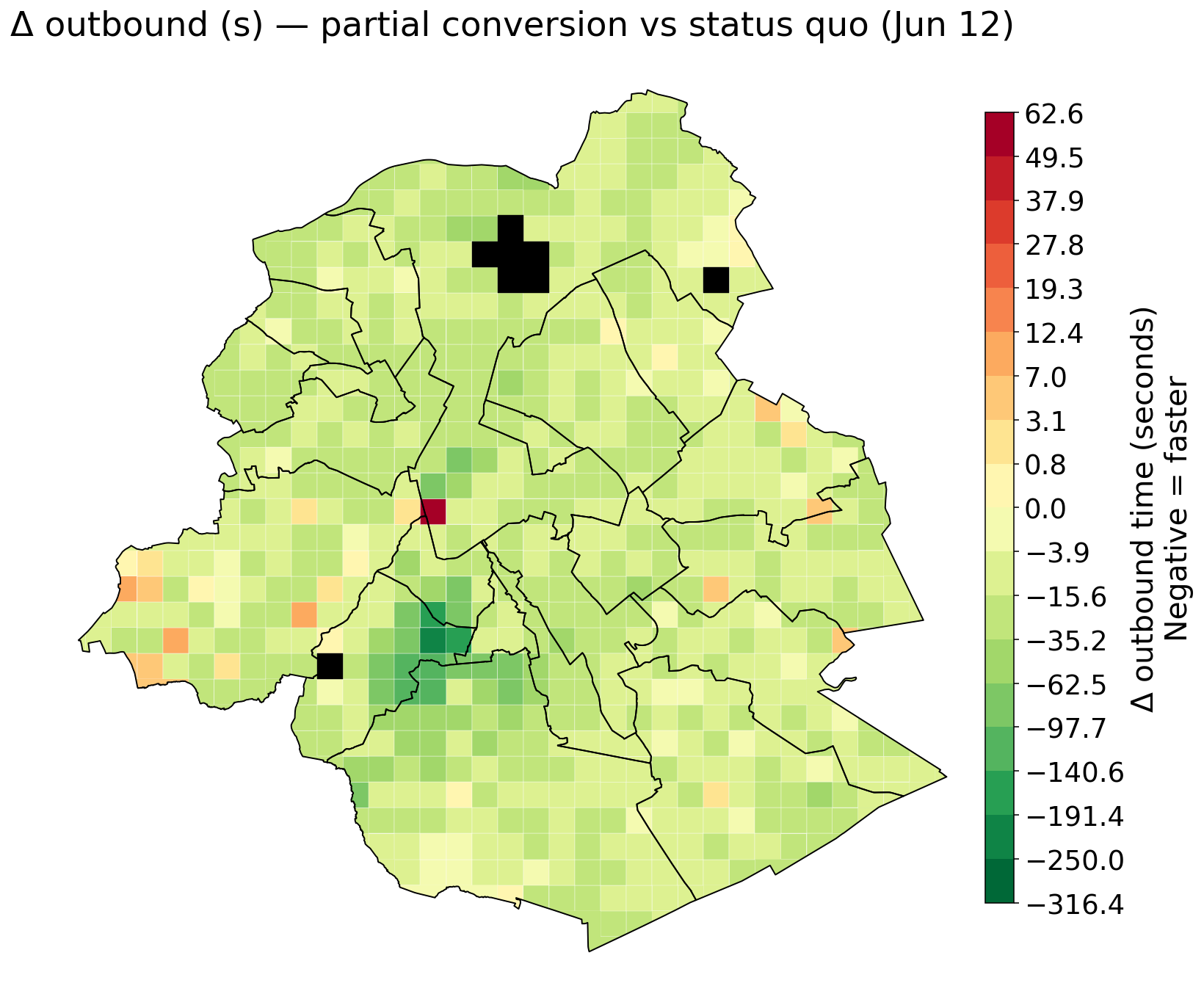}
    \caption{0612 — partial}
    \label{fig:map0612-partial}
  \end{subfigure}\hfill
  \begin{subfigure}[t]{0.32\linewidth}
    \centering
    \includegraphics[width=\linewidth]{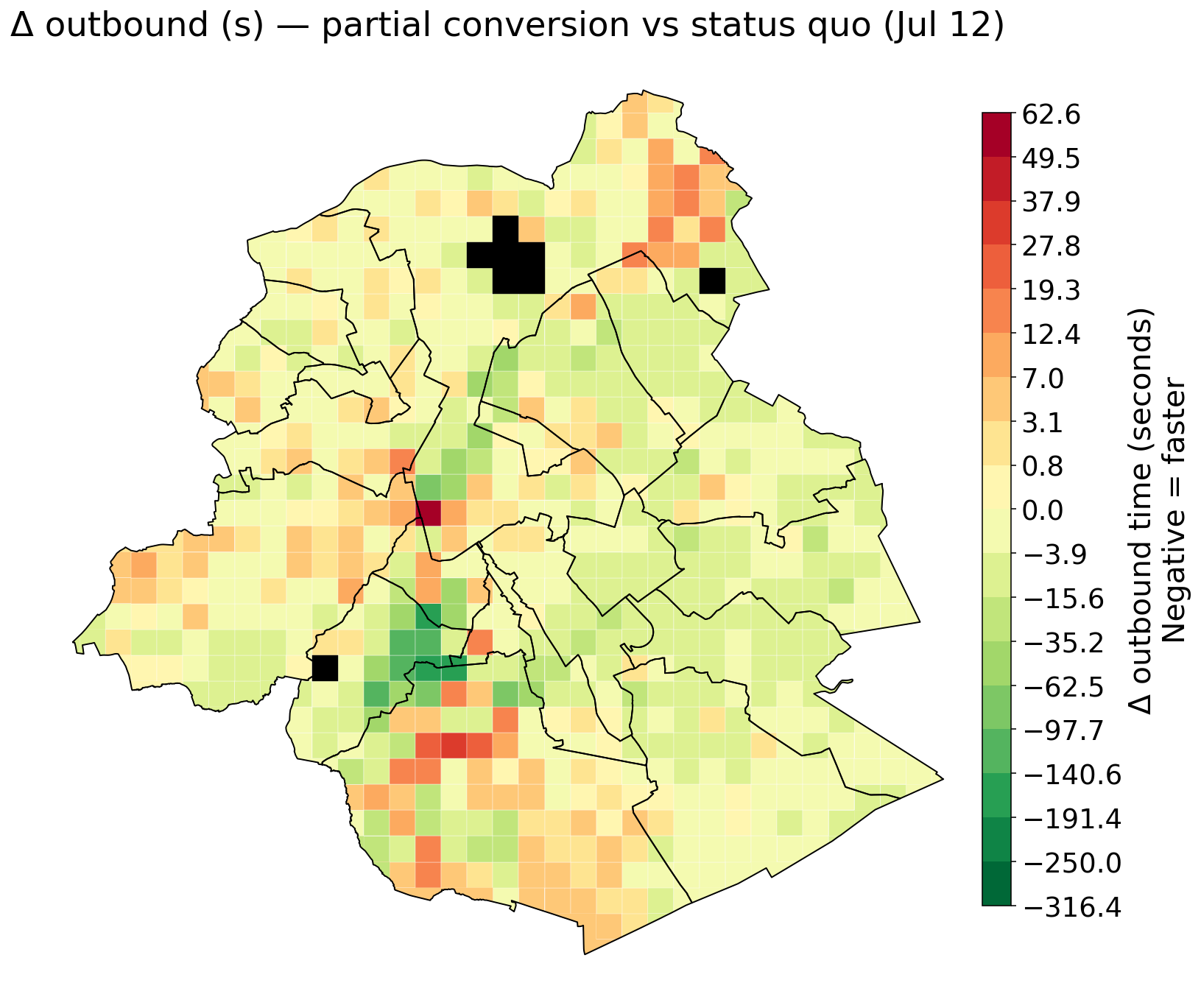}
    \caption{0712 — partial}
    \label{fig:map0712-partial}
  \end{subfigure}

  \vspace{0.75em}

  % Row 2
  \begin{subfigure}[t]{0.32\linewidth}
    \centering
    \includegraphics[width=\linewidth]{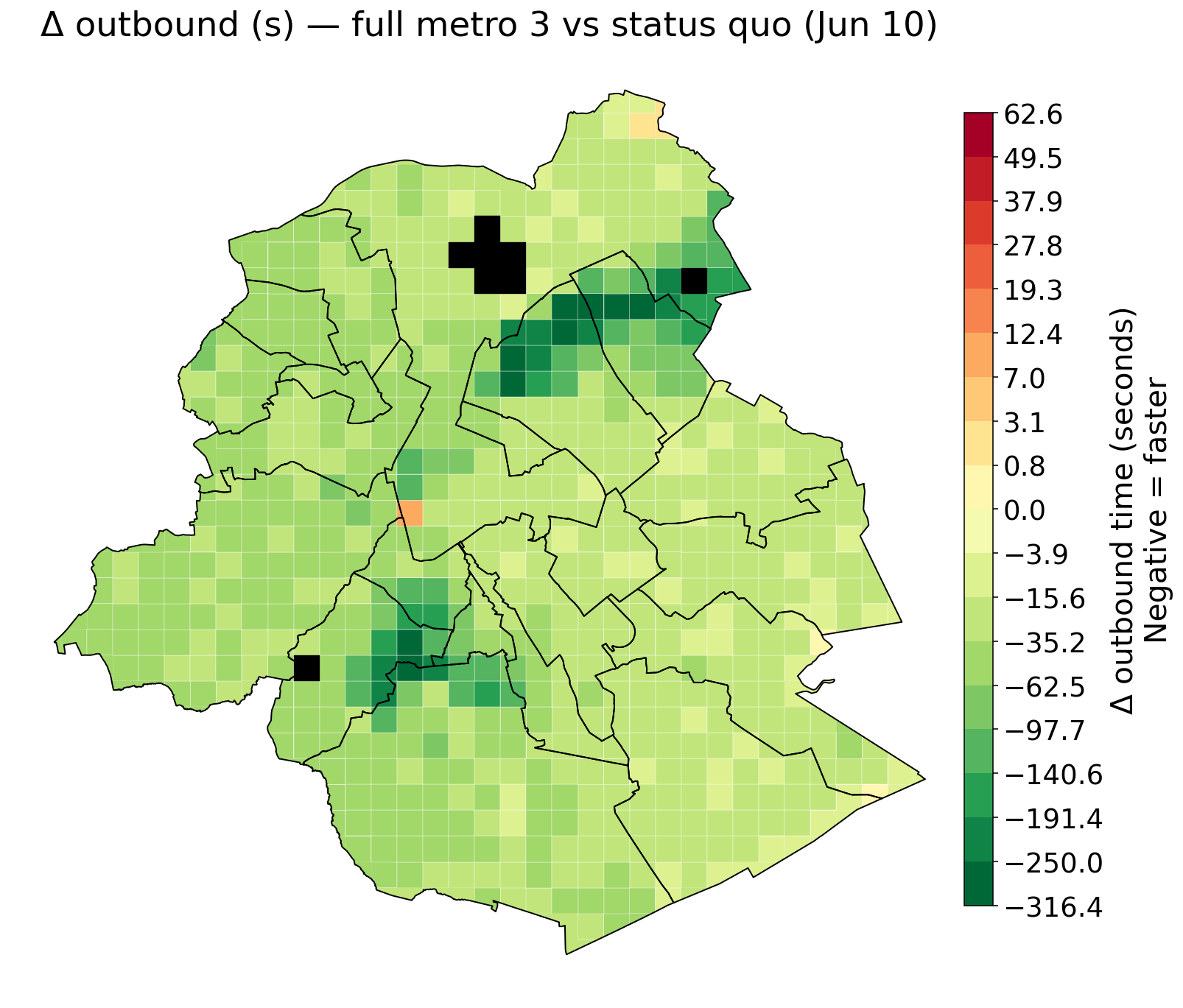}
    \caption{0610 — full}
    \label{fig:map0610-full}
  \end{subfigure}\hfill
  \begin{subfigure}[t]{0.32\linewidth}
    \centering
    \includegraphics[width=\linewidth]{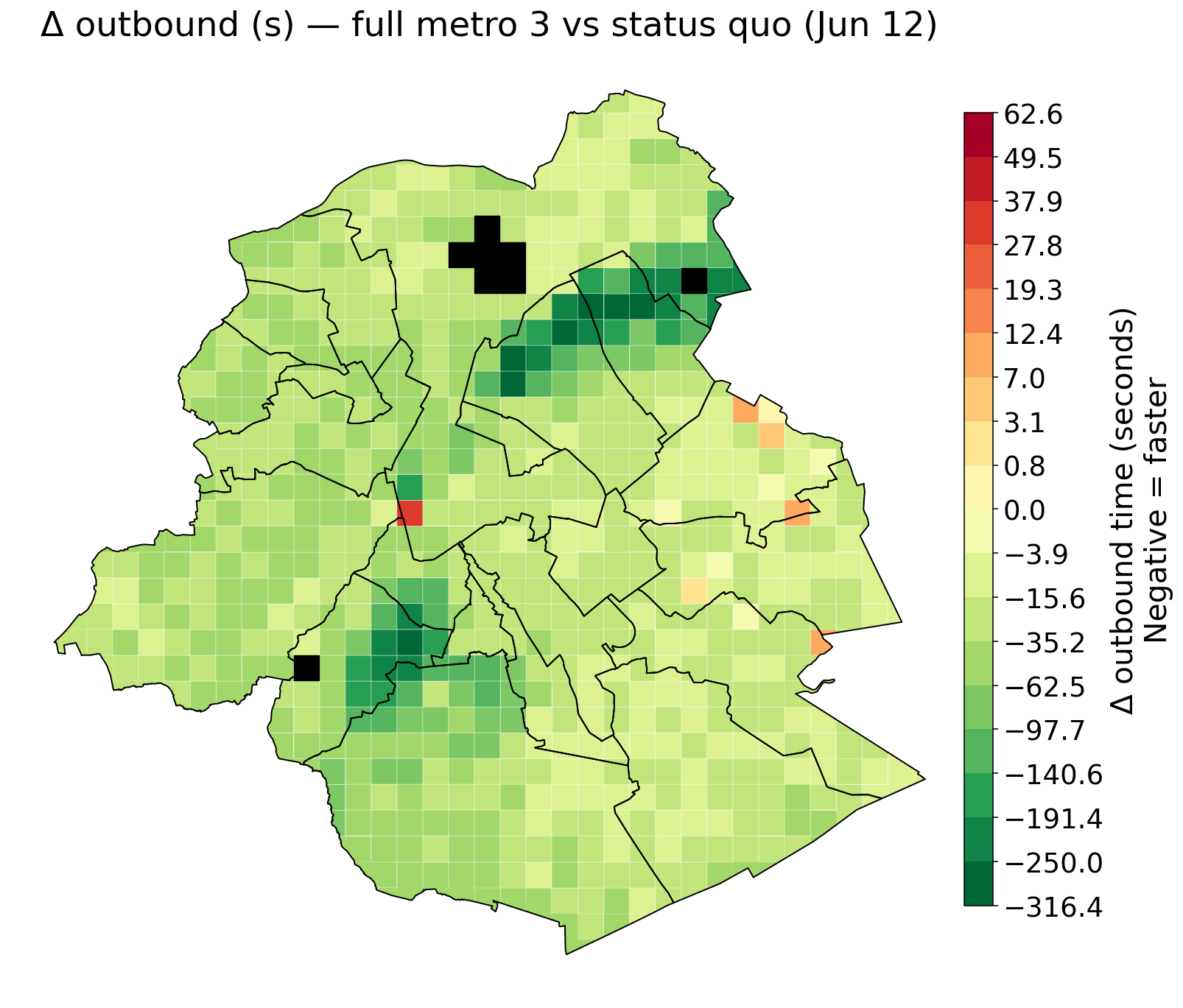}
    \caption{0612 — full}
    \label{fig:map0612-full}
  \end{subfigure}\hfill
  \begin{subfigure}[t]{0.32\linewidth}
    \centering
    \includegraphics[width=\linewidth]{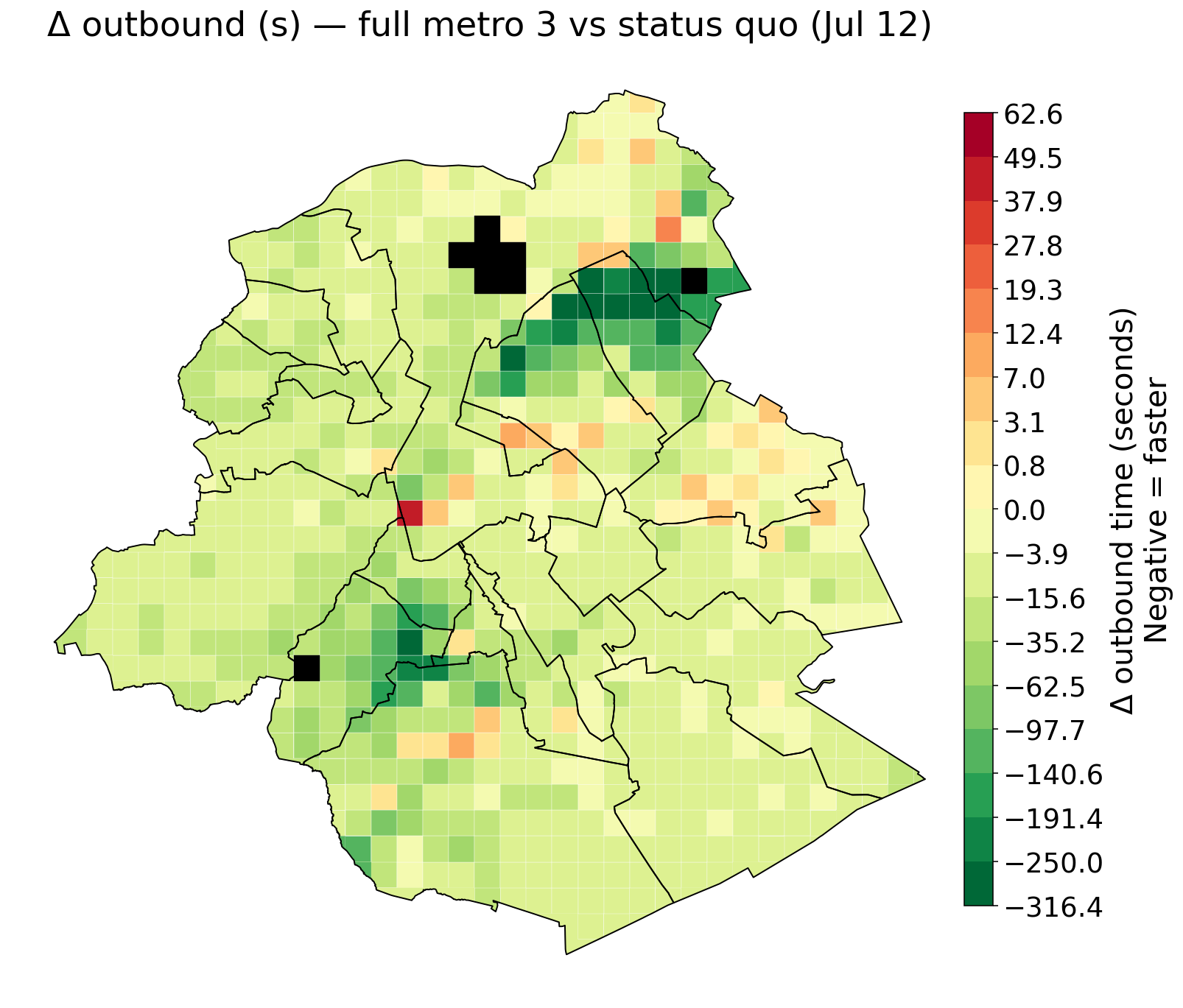}
    \caption{0712 — full}
    \label{fig:map0712-full}
  \end{subfigure}

  \caption{\textbf{Dated maps.} Daily snapshots by date (0610, 0612, 0712) and scenario (partial/full) of differences in means for partial conversion and full line.}
  \label{fig:maps-dated}
\end{figure}
\end{landscape}
\restoregeometry
\clearpage

\clearpage
\newgeometry{left=1cm,right=1cm,top=1cm,bottom=1cm} % temporary tighter margins
\begin{landscape}
\thispagestyle{empty}
\begin{figure}[p]
  \centering
  \begin{subfigure}[t]{0.49\linewidth}
    \centering
    \includegraphics[width=\linewidth]{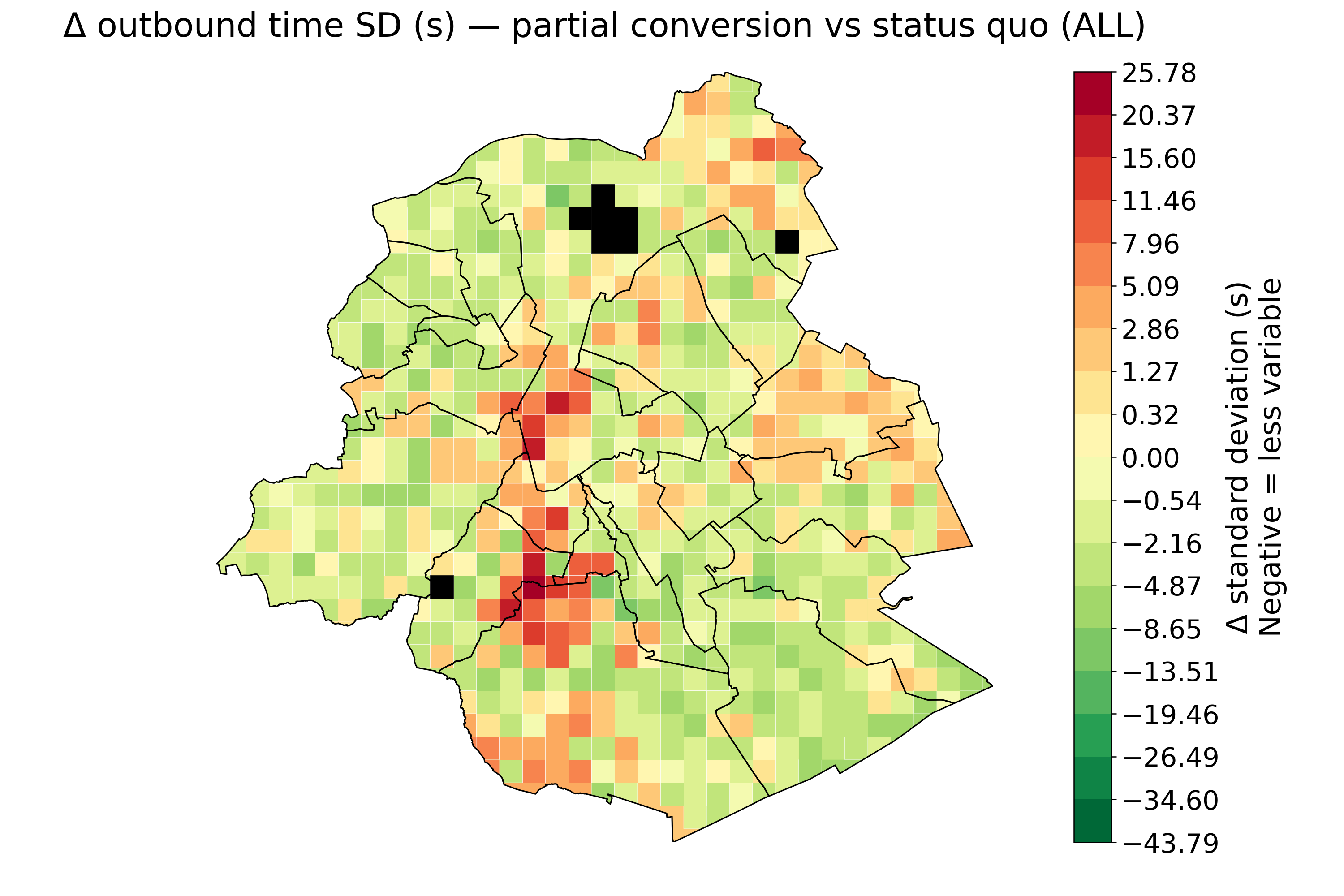}
    \caption{\textbf{All-days composite — partial conversion}}
    \label{fig:mapall-partial-sd}
  \end{subfigure}\hfill
  \begin{subfigure}[t]{0.49\linewidth}
    \centering
    \includegraphics[width=\linewidth]{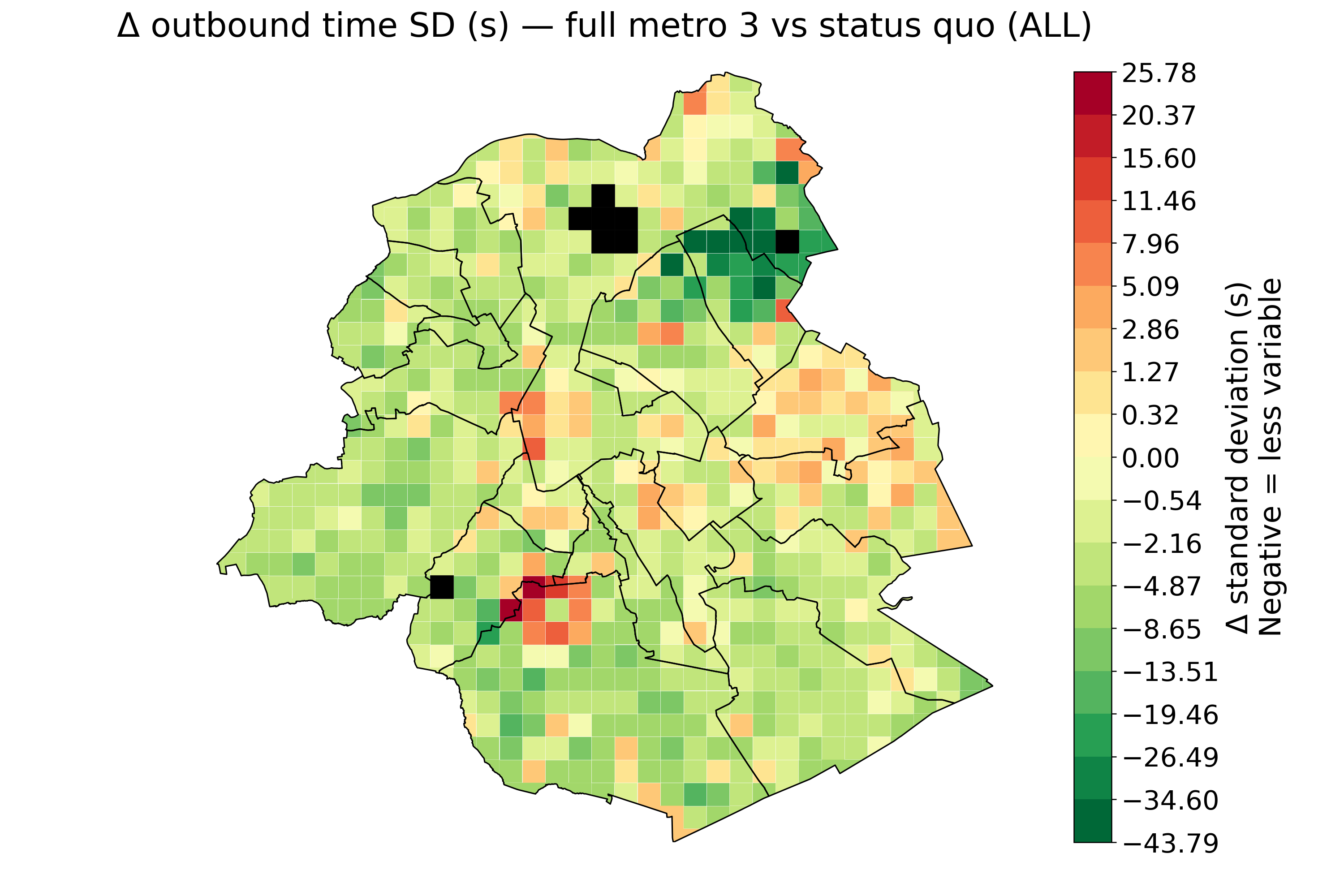}
    \caption{\textbf{All-days composite — full conversion}}
    \label{fig:mapall-full-sd}
  \end{subfigure}\hfill

  \caption{\textbf{Systemwide maps.} All-days composites for differences in standard deviation for partial conversion and full line.}
  \label{fig:maps-priority-sd}
\end{figure}
\end{landscape}
\restoregeometry
\clearpage

\subsection{Reliability and risk-adjusted accessibility metrics}
\label{sec:reliability_ce}
\paragraph{Per–OD, per–day reliability statistics}
For each scenario $s$, OD pair $(i,j)$, and day $d$, let the three canonical departure slices be $\mathcal{K}=\{-10,0,+10\}$ and denote by $T_{s,d,k}(i,j)$ the simulated travel time (s) at slice $k\in\mathcal{K}$. Define the empirical quantile

\[
Q_{p,s,d}(i,j)
\;=\;
\operatorname{Quantile}\!\big(\{\,T_{s,d,k}(i,j)\,:\,k\in\mathcal{K}\,\},\,p\big).
\]
We report
\begin{align}
\operatorname{median}_{s,d}(i,j) \;&=\; Q_{0.50,s,d}(i,j),\\
\operatorname{IQR}_{s,d}(i,j) \;&=\; Q_{0.75,s,d}(i,j)-Q_{0.25,s,d}(i,j),\\
\operatorname{RBI}^{\text{abs}}_{s,d}(i,j) \;&=\; Q_{0.95,s,d}(i,j)-Q_{0.50,s,d}(i,j),\\
\operatorname{RBI}^{\text{rel}}_{s,d}(i,j) \;&=\; \frac{Q_{0.95,s,d}(i,j)}{Q_{0.50,s,d}(i,j)}-1.
\end{align}

Here $\operatorname{RBI}^{\text{abs}}$ is an absolute ``reliability buffer'' (95th–50th) widely used in reliability assessment~\citep{chen2004systematic}, while $\operatorname{RBI}^{\text{rel}}$ is the median-based Buffer Index (unitless; $\times 100$ for \%), a robust variant analogous to the mean-based Buffer Index  \citep{lyman2008using}.

\paragraph{Risk-adjusted (entropic) certainty equivalent} 
Many OD pairs exhibit skew with occasional long delays; we therefore seek a single summary that rewards lower typical times while penalising tail risk, echoing the time-average perspective which highlights the structural importance of reliability alongside speed~\citep{Vanhoyweghen2025Ergodicity}. The entropic certainty equivalent answers: \emph{“What certain travel time would make a traveller, who values both travel speed and reliability, indifferent to the observed distribution?”} Lower is better; the gap $\mathbb{E}[T]-\mathrm{CE}_\rho[T]$ is the risk premium (in minutes) the traveller would give up to avoid variability.

To penalise tail risk we adopt the entropic certainty equivalent for travel time $T$:
\[
\mathrm{CE}_{\rho}[T]\;=\;-\frac{1}{\rho}\log\Big(\mathbb{E}\big[e^{-\rho T}\big]\Big),
\]
with $\rho>0$ calibrated by a “half-life’’ parameter. 
We fix the half-life at $20$ min after testing multiple values ($\rho = \frac{\ln 2}{20 \times 60~\text{s}} 
      = \frac{0.693147}{1200~\text{s}} 
      \approx 5.78 \times 10^{-4}~\text{s}^{-1}$); results were qualitatively stable across choices, and 20 min offers a clear interpretation: an extra 20 min halves the effective weight.

A second-order expansion shows
\[
\mathrm{CE}_{\rho}[T]\approx \mathbb{E}[T]-\tfrac{\rho}{2}\mathrm{Var}(T),
\]
i.e.\ the mean minus a variance penalty that grows with $\rho$~\citep{follmer2011stochastic}. Numerically, we use a log-sum-exp stabilised estimator:
\[
\widehat{\mathrm{CE}}_{\rho}=-\frac{1}{\rho}\Big(\log\frac{1}{L}\sum_{\ell=1}^{L}e^{-\rho T_{s,d,\ell}} \Big).
\]

\emph{Illustration (in minutes, with $\rho=\ln 2 / 20$):} 
consider a travel time $T$ that is $30$ with probability $0.9$ and $70$ with probability $0.1$. 
The mean is $\mathbb{E}[T]=34$, while the certainty equivalent is 
\[
\mathrm{CE}_\rho[T] \;\approx\; 32.25,
\]
so the implied \emph{risk premium} is $\mathbb{E}[T]-\mathrm{CE}_\rho[T]\approx 1.75$ minutes. 
This reflects the fact that travellers prefer a sure $32.25$ minutes to facing the $90\%$–$10\%$ gamble. 
By contrast, for a degenerate case $T\equiv 34$, the certainty equivalent coincides with the mean: 
$\mathrm{CE}_\rho[T]=34$.

\paragraph{Scenario-level reporting}
Let \(\mathcal S\) denote the OD pairs (as defined earlier) and \(\mathcal D\) the service days. 
For \(M\in\{p_{50},\mathrm{IQR},\mathrm{RBI}^{\mathrm{abs}},\mathrm{RBI}^{\mathrm{rel}},\mathrm{CE}_\rho\}\), define
\[
\overline{M}_{s,d}=\frac{1}{|\mathcal S|}\sum_{(i,j)\in\mathcal S} M_{s,d}(i,j),\qquad
\Delta \overline{M}_{s,d}=\overline{M}_{s,d}-\overline{M}_{\text{base},d},
\]
and report
\[
\Delta \overline{M}_{s}=\frac{1}{|\mathcal D|}\sum_{d\in\mathcal D}\Delta \overline{M}_{s,d}.
\]

Interpretation: $\Delta \overline{p_{50}}_s<0$ implies faster typical trips; 
$\Delta \overline{\mathrm{RBI}}^{\mathrm{abs}}_s<0$ implies a smaller absolute buffer (seconds); 
$\Delta \overline{\mathrm{RBI}}^{\mathrm{rel}}_s<0$ implies a smaller median-based Buffer Index (more reliable in relative terms); 
$\Delta \overline{\mathrm{CE}}_s<0$ implies a lower risk-adjusted travel time for penalty $\rho$.

\clearpage
\newgeometry{left=1cm,right=1cm,top=1cm,bottom=1cm}
\begin{landscape}
\thispagestyle{empty}
\begin{figure}[p]
  \centering

  % Left column: Spesta (partial above full)
  \begin{subfigure}[t]{0.4\linewidth}
    \centering
    \includegraphics[width=\linewidth]{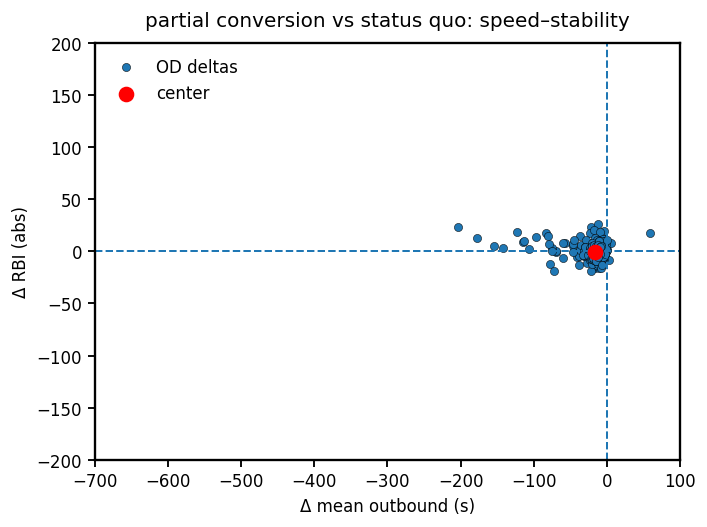}
    \vspace{0.75em}
    \includegraphics[width=\linewidth]{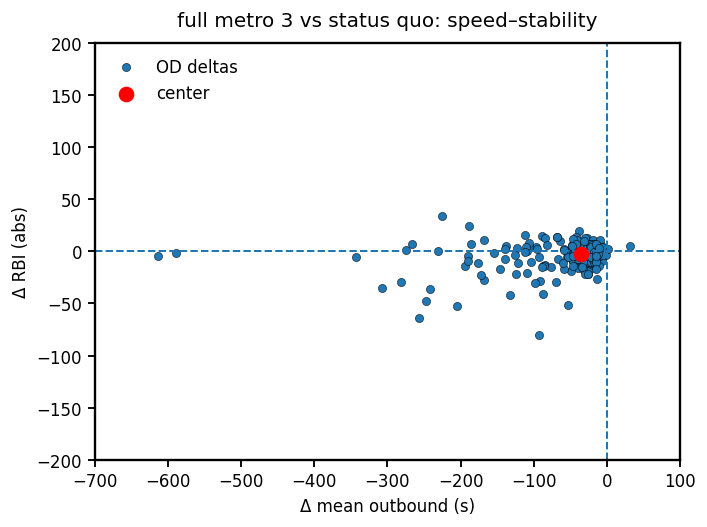}
  \end{subfigure}\hspace{0.02\linewidth}
  % Right column: Ramean (partial above full)
  \begin{subfigure}[t]{0.4\linewidth}
    \centering
    \includegraphics[width=\linewidth]{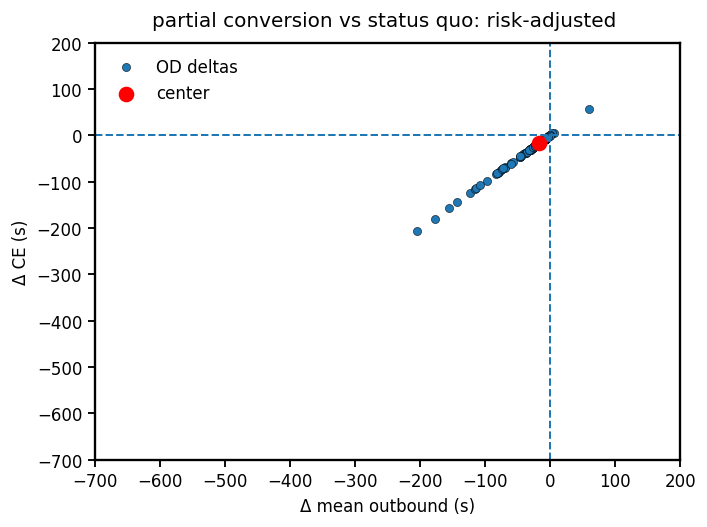}
    \vspace{0.75em}
    \includegraphics[width=\linewidth]{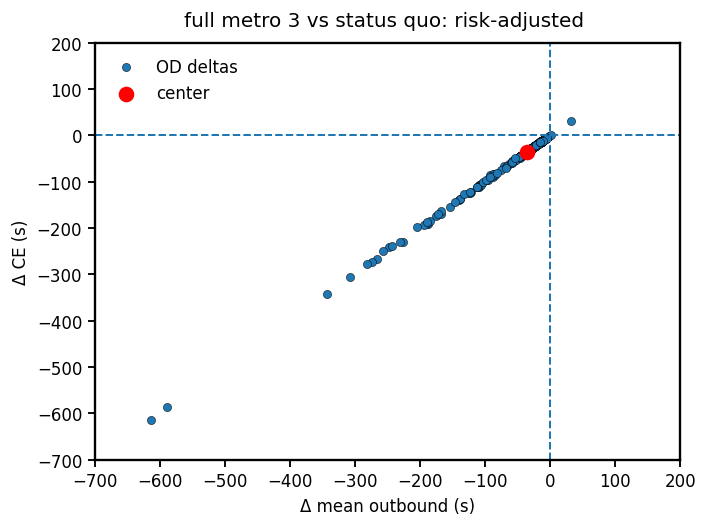}
  \end{subfigure}

 \caption{Mean effects across all days on OD pairs. Deltas are scenario $-$ status quo; the red dot is the cloud mean. Left panels: $\Delta$ mean outbound (s) vs.\ $\Delta\mathrm{RBI}_{\text{abs}}$ (s). Right panels: $\Delta$ mean outbound (s) vs.\ $\Delta\mathrm{CE}$ (s), with CE the certainty-equivalent time under exponential utility, $\rho=\ln 2/(20,\text{min})$. Dashed axes mark no change; left/down is better. Quadrants: $(-,-)$ win; $(-,+)$ faster/less stable; $(+,-)$ slower/more stable; $(+,+)$ dominated.}
  \label{fig:spesta-ramean}
\end{figure}
\end{landscape}
\restoregeometry
\clearpage

\begin{table}[htbp]
\centering
\scriptsize
\begin{adjustbox}{max width=\textwidth}
\begin{tabular}{l S S S S S}
\toprule
scenario 
& {$\Delta p_{50}$ (s)} 
& {$\Delta IQR$ (s)} 
& {$\Delta RBI_{\text{abs}}$ (s)} 
& {$\Delta RBI_{\text{rel}}$} 
& {$\Delta CE$ (s)} \\
\midrule
partial conversion 
& -16.61 & -0.61 & -0.43 & 0.018\%
 & -16.42 \\
\addlinespace
full metro 3 
& -36.34 & -3.70 & -3.01 & -0.023\%
 & -35.60 \\
\bottomrule
\end{tabular}
\end{adjustbox}
\caption{Averages over the three service days of robustness-aggregated OD time-change metrics. Negative values indicate improvements.}
\label{tab:robustness}
\end{table}

\section{Results}
\label{sec:results}

\subsection{System-wide time benefits and their distribution}

Table~\ref{tab:summary} summarises OD-level time changes relative to the status quo. The \emph{partial conversion} reduces mean OD time by \SI{-16.5}{\second} with a zero median; \SI{49.8}{\percent} of OD pairs improve and \SI{5.6}{\percent} improve by at least \SI{5}{\percent}. The \emph{full line} roughly doubles the mean effect, resulting in about half a minute gain for the average one-way commute (mean \SI{-36.1}{\second}, median \SI{-6.1}{\second}). The \emph{full line} increases the share improved to \SI{52.8}{\percent}, and lifts the share with \SI{\ge 5}{\percent} gains to \SI{9.4}{\percent} (and \SI{2.8}{\percent} with \SI{\ge 10}{\percent} gains).

Distributional diagnostics show that these averages hide considerable heterogeneity. Deciles of $\Delta T$ in Table~\ref{tab:percentiles} and Fig.~\ref{fig:quantiles} indicate broad left shifts up to about the 60th–70th percentile (improvements), with thin right tails (moderate slowdowns). Extreme improvements exist: the best \SI{0.1}{\percent} of OD pairs save \SIrange{-700}{-1700}{\second} (partial vs.\ full). The \emph{shift function} (Fig.~\ref{fig:shift}) makes this transparent: for the full line, the trough around $p\!\approx\!0.4$–$0.6$ reaches $\sim\SI{-50}{\second}$, versus $\sim\SI{-20}{\second}$ for the partial conversion. In other words, the bulk of trips get a small but systematic speed-up, larger under the full build.

The difference in empirical CDFs, $\Delta \widehat F(t)$ (Fig.~\ref{fig:decdf}), quantifies how many more OD pairs complete within a given time budget. Gains peak around $t\!\approx\!\SIrange{3300}{3800}{\second}$ : roughly +\SI{1.8}{pp} for the full line and +\SI{0.75}{pp} for the partial conversion. These are distribution-wide effects, not driven by a handful of OD pairs.

\subsection{Origin-level accessibility changes and robustness}

Per-origin outbound changes, $\Delta^{\mathrm{out}}$ (Fig.~\ref{fig:distros}), are centered below zero for all service days; the full line yields larger gains and slightly larger dispersion than the partial conversion. Robustness diagnostics using three departure instants per band (\S\ref{sec:methods}) show that the sign and magnitude of benefits are stable across the AM peak, PM peak, and Saturday; tails are heavier for the full line but remain predominantly on the benefit side.

% Figure description
Figures~\ref{fig:spesta-ramean} show origin-averaged deltas (scenario $-$ status quo) pooled over all service days.
Left panels: $\Delta$ median vs.\ $\Delta\mathrm{RBI}$ with $\mathrm{RBI}=p95-p50$ (smaller = more stable).
Right panels: $\Delta$ median vs.\ $\Delta\mathrm{CE}$ under exponential utility with a 20-min half-life (smaller = better risk-adjusted time).
Dashed axes mark parity; left/down is improvement. Quadrants: lower-left = win; upper-left = faster/less stable (or worse CE);
lower-right = slower/more stable (or better CE); upper-right = dominated. Other half-lives yield similar results as $\mathrm{CE}$ is dominated by the mean here. 

% 3–5 line interpretation
Partial conversion yields modest average time savings (a few–tens of seconds) with near-zero change in stability; CE gains are small and track the mean closely.
Full metro 3 shifts the cloud further left and delivers correspondingly larger CE improvements; stability is mostly neutral to slightly better, with some notable RBI drops.
Overall the mass of points sits in the “win” quadrant with only a few mild regressions, and the near-linear CE–mean relationship implies distributional changes are limited relative to the central-tendency shift.

Partial conversion.  
Speed–stability: points concentrate tightly around the origin, with most lying slightly left but above the $x$-axis ($\Delta$ median $\approx -60\ldots+30$ s; $\Delta\mathrm{RBI}\approx +5\ldots+25$ s). Interpretation: small median gains are paired with modestly worse stability, and true reliability improvements are rare.  
Risk-adjusted: scatter straddles the horizontal axis with magnitudes generally small ($\Delta\mathrm{CE}\approx -50\ldots+20$ s). Net: effects are modest and mixed—minor speedups tend to be offset by degraded stability, leaving CE nearly unchanged.

Full conversion.  
Speed–stability: strong leftward shift with most points below zero on both axes ($\Delta$ median typically $-120\ldots-30$ s, several reaching $-600$ s; $\Delta\mathrm{RBI}$ commonly $-10\ldots-60$ s). Interpretation: broad gains in both speed and stability; only a small minority trade stability for speed.  
Risk-adjusted: a pronounced down-left diagonal ($\Delta\mathrm{CE}$ closely tracking $\Delta$ median at $\sim$1:1), spanning $-50\ldots-600$ s. Nearly all origins improve CE substantially; dominated outcomes are rare.

We conclude that partial conversion yields modest gains in travel speed but risks deterioration in reliability, whereas full conversion delivers both substantial speed increases and consistent reliability improvements.

Aggregated reliability statistics (Table~\ref{tab:robustness}) confirm this: full-line averages are $\Delta p50=\SI{-36.3}{\second}$, $\Delta\mathrm{RBI}=\SI{-3.0}{\second}$, $\Delta\mathrm{CE}=\SI{-35.6}{\second}$; partial conversion effects are smaller in the same direction.

\subsection{Where the gains are: spatial pattern}

All-days composite maps (Fig.~\ref{fig:maps-priority}, Fig.~\ref{fig:maps-priority-sd}) and day-specific mosaics (Fig.~\ref{fig:maps-dated}) show a consistent geography on the OD pairs:

\begin{itemize}\setlength\itemsep{0.35em}
\item \textbf{Partial conversion.} Benefits concentrate along and south of the new line, with the strongest outbound improvements in \emph{Forest} and \emph{Saint-Gilles}, extending into northern \emph{Uccle} (see \ref{app:communes}, \autoref{fig:brussels-communes} for the location of communes). Elsewhere, gains are light and scattered or near zero.
\item \textbf{Full line.} The same southern core improves, and benefits propagate north–east into \emph{northern Schaerbeek} and \emph{Evere}, and north of \emph{Saint-Josse-ten-Noode} and the \emph{Ville de Bruxelles}, yielding a larger contiguous centre–north swath of improvements.
\end{itemize}

Per-day mosaics are more scattered than the composites, and the standard-deviation map (Fig.~\ref{fig:maps-priority-sd}) makes the heterogeneity explicit: under partial conversion, pockets of deterioration appear in \emph{Ixelles}, northern \emph{Uccle}, and parts of \emph{Forest}; under the full line, residual negatives persist north of \emph{Uccle} and in parts of \emph{Forest} but are smaller, while large positive effects concentrate north of the \emph{Ville de Bruxelles} and across \emph{Evere}. A structural downside of partial conversion might be the absence of a dedicated depot: rolling stock must be positioned from elsewhere each day, constraining staging and turnbacks and capping the number of trainsets in service.

Black tiles denote unreachable grid cells (six in Laeken Park, one within NATO grounds, one in the track interstice near Bruxelles-Midi) and have no analytical weight. The isolated \textit{red tile near Lemonnier} should \emph{not} be interpreted as a localised harm from the project: it is a data artefact of the 500\,m centroid being unusually close to the current tram stops (Trams~4 and 10) and further from the planned metro entrance. Because the tram services are replaced by the metro in our scenarios, the nearest-stop access distance increases for that particular centroid.

\subsection{What changes with a full build vs.\ a partial conversion}

Three points emerge from the head-to-head comparison:

\begin{enumerate}\setlength\itemsep{0.35em}
\item \textbf{Magnitude:} the full line roughly doubles the average benefit (mean \SI{-36}{\second} vs.\ \SI{-17}{\second}) and shifts the median below zero (Table~\ref{tab:summary}); its shift function trough is more than twice as deep (Fig.~\ref{fig:shift}).
\item \textbf{Breadth:} more OD pairs meet common time budgets (Fig.~\ref{fig:decdf}) and a larger share of origins experience improvements (Fig.~\ref{fig:distros}); spatially, benefits extend to the North-East (Fig.~\ref{fig:mapall-full}).
\item \textbf{Stability:} the full line’s gains do not come at the cost of higher variability; if anything, RBI declines slightly (Fig.~ref{fig:spesta-ramean}, Table~\ref{tab:robustness}).
\end{enumerate}

\subsection{Key quantitative takeaways}

\vspace{-0.35em}
\begin{itemize}\setlength\itemsep{0.35em}
\item Network-wide averages are modest but systematic: \SI{-16.5}{\second} (partial) and \SI{-36.1}{\second} (full) mean OD savings; medians $0$ and \SI{-6.1}{\second}. 
\item Around half of OD pairs improve; \SI{9.4}{\percent} (full) achieve $\ge\SI{5}{\percent}$ savings (Table~\ref{tab:summary}).
\item The biggest gains occur in Forest and Saint-Gilles (both scenarios), extending to northern Schaerbeek and Evere under the full scenario (Figs.~\ref{fig:maps-priority}–\ref{fig:maps-dated}).
\item Robustness checks across $t\!-\!10,\,t,\,t\!+\!10$ and three day types preserve all qualitative findings; risk-adjusted improvements track median improvements closely (Fig.~\ref{fig:spesta-ramean}).
\item Apparent local slowdowns at the single red tile near Lemonnier are a \emph{nearest-stop artefact} due to the grid centroid being nearer the existing tram stops than the future metro entrance; surrounding tiles follow the corridor-wide pattern of gains.
\end{itemize}

\section{Implications and limitations}
\label{sec:implications}

\paragraph{Time-savings ledger (lower bound, not a decision rule)}
Using the citywide mean gain of $16.5$–$36.1$\,s per trip and $400$–$500$ million trips/year, even with $\mathrm{VOT}$= \euro15/h the time-only benefit is just \euro60–\euro75\,m/year. Capitalised at 4\% over 40 years ($\mathrm{CRF}=0.0505$), this supports at most \euro1.5\,bn before O\&M and \euro0.3–\euro0.7 \,bn after plausible incremental O\&M (\euro40–\euro60\,m/year). Time savings alone therefore cannot carry a multi-billion project. We treat them as a lower bound and focus the appraisal on network option value, reliability and decarbonisation, and the distributional gains documented above.\footnote{Formulae: $B=|\overline{\Delta t}|\,Q\,\mathrm{VOT}/3600$; $\mathrm{CAPEX}_\star=(B-\mathrm{O\&M})/\mathrm{CRF}$ with $\mathrm{CRF}=r/(1-(1+r)^{-n})=0.0505$ for $r{=}0.04$, $n{=}40$. Full parameter ranges and sensitivity in App.~\ref{app:monetisation}.}

Against multi-billion capital outlays, this suggests that short-term time savings alone are unlikely to justify Metro~3. Rather, the case must rest on wider system benefits such as network option value, reliability, decarbonisation, and the distributional gains we document. In this broader perspective, the project could also alleviate pressure on the \emph{SNCB/NMBS} network by reducing reliance on the saturated North–South axis through Brussels, a persistent bottleneck in the national rail system~\citep{te2018new}. At the same time, it creates the opportunity for a well-connected park-and-ride facility in the city’s northern periphery, an element currently missing from Brussels’ transport offer, which would further enhance multimodal integration and accessibility. Of course, additional network effects or alternative interventions may also be relevant, and a full appraisal should consider these alongside the impacts we highlight.

\subsection{Policy interpretation: why a time-only CBA misses the point}

Three considerations make a conventional short-horizon, time-only CBA uninformative in this case.

\paragraph{(i) Network option value and path dependence}
Metro~3 is a \emph{structural} intervention: it creates a high-capacity backbone that future surface restructurings and radial/diagonal links can hinge on. The option value of a grade-separated north–south trunk in Brussels (long identified as a binding network bottleneck) cannot be proxied by first-year time savings. It is an investment in \emph{state space}: it expands what the network can become (and how it can be defended operationally) under growth, diversion, incident, and maintenance regimes.

\paragraph{(ii) Externalities and non-marginal mode shift}
Our schedule-based, person-time results do not monetize \emph{external} effects. At network scale, higher metro mode share plausibly reduces GHG/PM/NOx, crash risk, and surface congestion, enables bus fleet repurposing, and supports station-area intensification; countervailing, place-based costs include construction-stage disruption, operational noise/vibration, platform/corridor crowding, risks of station-area social disorder (including visible homelessness), and local parking spillovers. Because the intervention induces \emph{non-marginal} mode shift, externalities are non-linear: network benefits grow with long-run mode shares while certain disamenities scale with exposure at specific stations.

\paragraph{(iii) Distribution and resilience as first-order policy goals}
The largest gains locate in \emph{Forest} and \emph{Saint-Gilles} for both scenarios, and extend into \emph{northern Schaerbeek} and \emph{Evere}  for the full line (Figs.~\ref{fig:maps-priority}–\ref{fig:maps-dated}). These encompass densely populated, persistently disadvantaged neighbourhoods \citep{IBSA2024MiniBru} . If the policy objective includes reducing \emph{spatial inequality in access}, an equity-weighted welfare function would overweight these benefits and tilt appraisal in favour of the full build. Moreover, our robustness diagnostics show that improvements are not bought at the cost of instability; reliability metrics improve (Table~\ref{tab:robustness}), which matters disproportionately to lower-income travellers facing tighter temporal constraints.

\subsection{Decision-relevant implications}

\begin{itemize}\setlength\itemsep{0.35em}
\item \textbf{Partial vs full.} The full line roughly doubles time benefits and extends them into the North-East, reaching additional vulnerable areas, without degrading reliability. If the decision shifts toward phasing, the analysis suggests that stopping at a partial conversion sacrifices a large share of distributionally valuable gains.

\item \textbf{Transparency.} The absence of any official cost–benefit analysis underscores the need for open, reproducible evidence. Publishing ``future GTFS’’ feeds and the scenario-building logic would enable independent replication and allow policy to evolve iteratively: tram and bus restructurings could be stress-tested transparently alongside the trunk alignment, rather than debated in the dark.

\end{itemize}

\subsection{Limitations (and why they matter)}

\paragraph{Coarse spatial sampling}
We use a $500{\times}500$\,m grid (639 effective centroids) as a tractability–coverage compromise. This smooths within-neighbourhood variation in station access/egress and can create artefacts where \newline centroid–entrance geometry changes between scenarios (the Lemonnier dot). A denser lattice and/or network-constrained access modelling would sharpen micro-scale heterogeneity and reduce geometry-driven anomalies.

\paragraph{Scenario information scarcity}
There is almost no authoritative, stable, public technical documentation at stop-by-stop resolution: station footprints, entrance locations, in-complex transfer links, dwell policies, depot/turnback constraints, and surface restructurings are pieced together from dated releases and press communications. Our GTFS scenario feeds are therefore \emph{internally coherent counterfactuals}, not official plans. Exact implementation choices could move neighbourhood-scale results by dozens of seconds; this is a tractable uncertainty, but it requires the promoter to publish definitive operational data.

\paragraph{No demand model, no boarding loads}
We report \emph{generalised time} only. Without a demand model and APC/OD estimates, we cannot weight OD benefits by realised passenger flows; this is precisely the step that would allow for a more credible welfare estimate. Similarly, without capacity and crowding we cannot capture queueing delays, denied boardings, or the relief on overloaded surface lines, effects that are often decisive for welfare. \emph{In this setting we implicitly treat all OD pairs and paths as equally important.} This uniform weighting ignores heterogeneity induced by workplace concentration, rail interchanges, and population density. Introducing OD weights (from APC/smartcard OD matrices or defensible gravity priors) would reweight accessibility toward the north–south spine—Belgium’s busiest axis. So, all else equal, the non-baseline scenarios would receive larger welfare gains relative if this was incorporated, in this sense our estimation can be viewed as a lower bound. This is an expected direction-of-effect; the magnitude depends on the adopted weights and time-of-day mix.

\paragraph{Temporal sampling and reliability}
Our robustness is intentionally minimalistic (three instants in three bands), inspired by ergodicity arguments: avoid single-snapshot bias. But we still treat schedules deterministically and do not stochastically model disruptions, blockages, or headway variance propagation. A full reliability appraisal would embed a stochastic service-control model and compute risk-sensitive accessibility over a richer temporal ensemble.

\paragraph{Land-use feedbacks and dynamic network effects}
The appraisal horizon that matters for a metro is \emph{decades}, not the first operating year. Station area densification, relocation of activity patterns, and induced demand typically magnify accessibility impacts over time and change the spatial distribution of benefits. Conversely, governance and cost risks can delay or truncate phases; a rigorous option-value treatment would place probabilities over these paths and price irreversibility.

\paragraph{Computational tractability}
The analysis required on the order of two CPU–years of computation. This shows the workflow is operating at the edge of the feasible computational frontier. Alternative modelling choices, finer lattices, network-constrained access, capacity and demand integration, would sharply increase runtime. We are therefore aware that other choices are possible and in some respects desirable, but their computational burden makes them currently impractical at metropolitan scale.

\subsection{What to conclude and what not to}

On a narrow, short-term \emph{time-savings} ledger, Metro~3 is \emph{probably not worth it}. The back-of-the-envelope monetisation above puts the full-line benefit at
\newline 
\(\approx\)\euro75.2\,m/year; with a 40-year horizon and 4\% discount rate this justifies \(\sim\)\euro1.5\,bn of capital \emph{before} O\&M and only \(\sim\)\euro0.3–0.7\,bn once realistic O\&M is included. In other words, time alone cannot carry a multi-billion project.

But that is not the right decision criterion. The question is whether Brussels needs a resilient, grade-separated north–south backbone to (i) enable a coherent surface network redesign, (ii) de-risk climate and mode-share goals, and (iii) reduce structural accessibility deficits in disadvantaged neighbourhoods. Our results speak directly to (iii) largest gains in Forest, Saint-Gilles, and, under the full line, into northern Schaerbeek and Evere, and they show these gains are \emph{robust} across departure-time perturbations and do not degrade reliability. Together, these are the kinds of \emph{distributional} and \emph{resilience} benefits a narrow CBA misses but policy cares about.

\textit{Position in the literature.} Substantively, we provide the first transparent, reproducible accessibility review of Metro~3. Methodologically, we contribute a generalisable pipeline for GTFS-based \emph{counterfactual} network evaluation under project uncertainty, with an explicit temporal-robustness design inspired by ergodicity economics (multi-instant, multi-day sampling) and risk-adjusted diagnostics. This combination is the paper’s core value: it produces decision relevant evidence where a full welfare CBA is presently infeasible, while advancing good practice for ex-ante accessibility analysis in transport geography.

\section*{Acknowledgments}
The resources and services used in this work were provided by the VSC (Flemish 
Supercomputer Center), funded by the Research Foundation -- Flanders (FWO) 
and the Flemish Government.
\section*{Declaration of generative AI and AI-assisted technologies in the writing process}
During the preparation of this work the author(s) used GPT-5 in order to assist with the drafting and structuring of the manuscript. After using this tool, the author(s) reviewed and edited the content as needed and take full responsibility for the content of the published article.

%\bibliographystyle{elsarticle-num} 
%\bibliography{doc/bib}



%\begin{thebibliography}{00}

%% For numbered reference style

%\bibitem{lamport94}
%  \textit{\LaTeX: a document preparation system},
%  Addison Wesley, Massachusetts,

%\end{thebibliography}

\newpage
\appendix
\section{Back-of-the-envelope monetisation: formulae and sensitivity}
\label{app:monetisation}

\subsection{Set-up and identities}
We monetise time savings as
\begin{align}
B \;=\; |\overline{\Delta t}|\times Q_{\text{trips/yr}}\times \frac{\mathrm{VOT}}{3600}
\quad\text{(\euro \, per year)}.
\end{align}
Annual benefits $B$ capitalise to breakeven CAPEX via the capital-recovery factor:
\begin{align}
\mathrm{CRF}(r,n) \;=\; \frac{r}{1 - (1{+}r)^{-n}},
\qquad
\mathrm{CAPEX}_\star \;=\; \frac{B-\mathrm{O\&M}}{\mathrm{CRF}}.
\end{align}
Everything is linear in $|\overline{\Delta t}|$, $Q$, and VOT; O\&M shifts CAPEX one-for-one divided by the CRF.

We use the paper’s measured mean time gains (partial: $16.5$\,s; full: $36.1$\,s), representative trip volumes $Q\in\{400,500\}\times 10^6$/yr, and VOT $\in\{12,15\}$\,\euro \,/h.

\begin{table}[htbp]
\centering
\scriptsize
\begin{adjustbox}{max width=\textwidth}
\begin{tabular}{l S[table-format=2.1] S[table-format=3.0] S[table-format=2.0] S[table-format=3.1]}
\toprule
Scenario & {$|\overline{\Delta t}|$ (s)} & {$Q$ (m trips/yr)} & {VOT (\euro \,/h)} & {$B$ (m\euro \,/yr)} \\
\midrule
Partial & 16.5 & 400 & 12 & 22.0 \\
Partial & 16.5 & 400 & 15 & 27.5 \\
Partial & 16.5 & 500 & 12 & 27.5 \\
Partial & 16.5 & 500 & 15 & 34.4 \\
\addlinespace
Full    & 36.1 & 400 & 12 & 48.1 \\
Full    & 36.1 & 400 & 15 & 60.2 \\
Full    & 36.1 & 500 & 12 & 60.2 \\
Full    & 36.1 & 500 & 15 & 75.2 \\
\bottomrule
\end{tabular}
\end{adjustbox}
\caption{Annual time-only benefits $B$ under the combinations used or cited in the text.}
\label{tab:Bpoint}
\end{table}

\begin{table}[htbp]
\centering
\scriptsize
\begin{tabular}{l S[table-format=1.5] S[table-format=1.5] S[table-format=1.5]}
\toprule
 & {\(n{=}30\)} & {\(n{=}40\)} & {\(n{=}50\)} \\
\midrule
\(r{=}3\%\) & 0.051019 & 0.043262 & 0.038865 \\
\(r{=}4\%\) & 0.057830 & 0.050523 & 0.046550 \\
\(r{=}5\%\) & 0.065051 & 0.058278 & 0.054777 \\
\bottomrule
\end{tabular}
\caption{Capital-recovery factor \(\mathrm{CRF}(r,n)\). Baseline in text: \(r{=}4\%\), \(n{=}40\Rightarrow \mathrm{CRF}=0.050523\).}
\label{tab:crf}
\end{table}

\subsection{From annual benefits to breakeven CAPEX}
Illustrative mapping from time-only annual benefit \(B\) to \(\mathrm{CAPEX}_\star\) for three discount rates (40-year horizon) and O\&M bands. Units: billion euros.

\begin{table}[htbp]
\centering
\scriptsize
\begin{adjustbox}{max width=\textwidth}
\begin{tabular}{l l S[table-format=1.1] S[table-format=1.1] S[table-format=1.1]}
\toprule
CRF case & \(B\) (m€/yr) & {O\&M=\(0\)} & {O\&M=\(40\)} & {O\&M=\(60\)} \\
\midrule
\(r{=}3\%\), \(n{=}40\) & 60.2 & 1.392 & 0.467 & 0.005 \\
\(r{=}3\%\), \(n{=}40\) & 75.2 & 1.738 & 0.814 & 0.351 \\
\addlinespace
\(r{=}4\%\), \(n{=}40\) & 60.2 & 1.192 & 0.400 & 0.004 \\
\(r{=}4\%\), \(n{=}40\) & 75.2 & 1.488 & 0.697 & 0.301 \\
\addlinespace
\(r{=}5\%\), \(n{=}40\) & 60.2 & 1.033 & 0.347 & 0.003 \\
\(r{=}5\%\), \(n{=}40\) & 75.2 & 1.290 & 0.604 & 0.261 \\
\bottomrule
\end{tabular}
\end{adjustbox}
\caption{Breakeven \(\mathrm{CAPEX}_\star=(B{-}\mathrm{O\&M})/\mathrm{CRF}\). The baseline row \(r{=}4\%,n{=}40\), \(B{=}75.2\) reproduces \(\sim\)\euro1.49\,bn before O\&M and \(\sim\)\euro0.70–0.30\,bn after \euro40–60\,m/yr O\&M.}
\label{tab:capexmap}
\end{table}

% Place this after \appendix in your file
% After \appendix
\clearpage
\section{Brussels communes map}\label{app:communes}

\begin{figure}[htbp]
\centering
\includegraphics[width=0.6\linewidth]{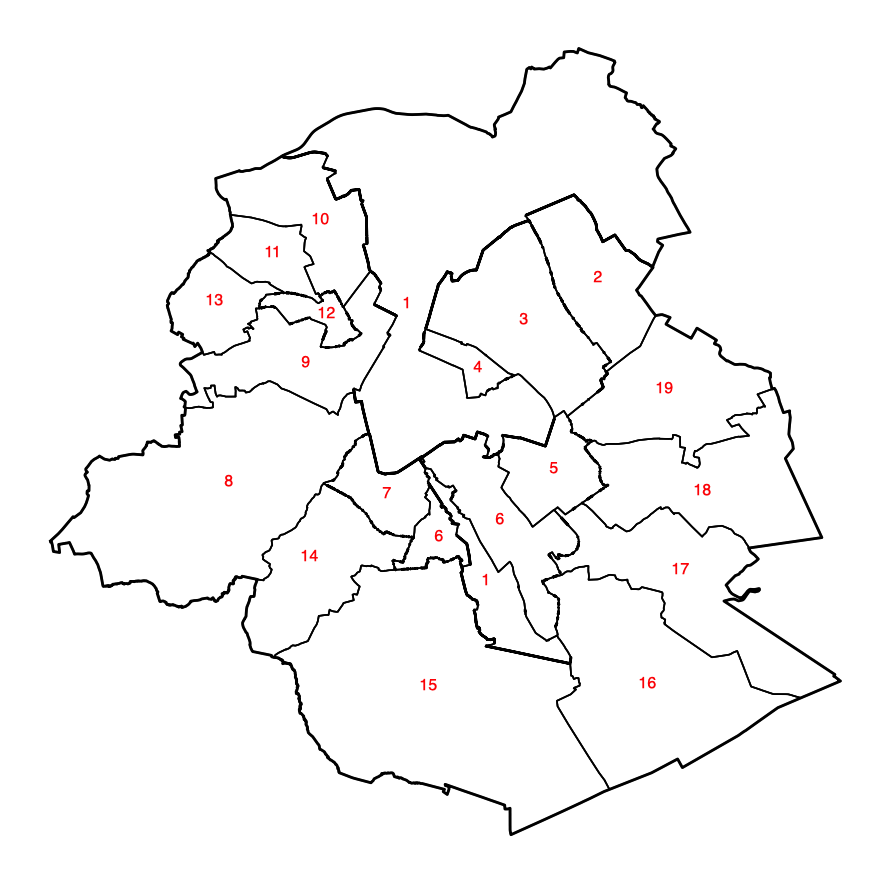}

\par\bigskip
\begin{minipage}{0.72\linewidth}
\footnotesize
\setlength{\columnsep}{1em}%
\begin{multicols}{2}
\begin{enumerate}\setlength\itemsep{0.2em}
    \item Ville de Bruxelles
    \item Evere
    \item Schaerbeek
    \item Saint-Josse-ten-Noode
    \item Etterbeek
    \item Ixelles
    \item Saint-Gilles
    \item Anderlecht
    \item Molenbeek-Saint-Jean
    \item Jette
    \item Ganshoren
    \item Koekelberg
    \item Berchem-Sainte-Agathe
    \item Forest
    \item Uccle
    \item Watermael-Boitsfort
    \item Auderghem
    \item Woluwe-Saint-Pierre
    \item Woluwe-Saint-Lambert
\end{enumerate}
\end{multicols}
\end{minipage}
\caption{Brussels-Capital Region communes. Numbers correspond to labels on the map.}
\label{fig:brussels-communes}
\end{figure}

\end{document}